\documentclass[prx,reprint,amsmath,amssymb,floatfix,citeautoscript,superscriptaddress]{revtex4-1}
\usepackage{cancel}
\usepackage{amsmath}
\usepackage{physics}
\usepackage{graphicx}
\usepackage{amssymb}
\usepackage{caption, subcaption}
\usepackage{sidecap}
\usepackage{amsthm}
\usepackage{bm}
\usepackage{dcolumn}
\usepackage{braket}

\usepackage{ragged2e}
\usepackage{txfonts}

\usepackage{color}
\usepackage[usenames,dvipsnames]{xcolor}
\definecolor{myblue}{rgb}{0,0,1}
\usepackage[breaklinks=true,colorlinks=true,linkcolor=myblue,citecol
or=myblue]{hyperref}
\usepackage{array}
\usepackage{multirow}

\newcommand{\I}{\mathrm{i}}

\newcommand{\paren}[1]{\left( #1 \right)}
\newcommand{\OO}[1]{\mathcal{O}\paren{#1}}
\newcommand{\wb}[1]{\overline{#1}}
\newcommand{\sump}[2]{\sideset{}{'}\sum_{#1}^{#2}}
\newcommand{\sumplim}[2]{\sideset{}{'}\sum\limits_{#1}^{#2}}

\newcommand{\mat}{\mathrm{M}}

\newcommand{\beq}[1]{\begin{equation} #1 \end{equation}}

\newcommand{\GR}{G^R}
\newcommand{\GLM}{G^\rceil}
\newcommand{\GL}{G^<}

\newcolumntype{M}[1]{>{\centering\arraybackslash}m{#1}}

\begin{document}
\title{Low rank compression in the numerical solution of the nonequilibrium Dyson equation}

\author{Jason Kaye}
\affiliation{Center for Computational Mathematics, Flatiron Institute,
New York, New York 10010, USA}
\affiliation{Center for Computational Quantum Physics, Flatiron Institute,
New York, New York 10010, USA}
\author{Denis Gole\v z}
\affiliation{Center for Computational Quantum Physics, Flatiron Institute,
New York, New York 10010, USA}

\begin{abstract}
We propose a method to improve the computational and memory efficiency of numerical solvers for the nonequilibrium Dyson equation in the Keldysh formalism. It is based on the empirical observation that the nonequilibrium Green's functions and self energies arising in many problems of physical interest, discretized as matrices, have low rank off-diagonal blocks, and can therefore be compressed using a  hierarchical low rank data structure. We describe an efficient algorithm to build this compressed representation on the fly during the course of time stepping, and use the representation to reduce the
  cost of computing history integrals, which is the main computational
  bottleneck. For systems with the hierarchical low rank property, our
  method reduces the computational complexity of solving the
  nonequilibrium Dyson equation from cubic to near quadratic, and the
  memory complexity from quadratic to near linear. We demonstrate the full
  solver for the Falicov-Kimball model exposed to a rapid ramp and
  Floquet driving of system parameters, and are able to
  increase feasible propagation times substantially. We present examples with
  $262\,144$ time steps, which would require approximately five months
  of computing time and $2.2$ TB of memory using the direct time stepping method,
  but can be completed in just over a day on a laptop with
  less than $4$ GB of memory using our method.
  We also confirm the hierarchical low
  rank property for the driven Hubbard model in the weak coupling regime
  within the GW approximation, and in the strong coupling regime
  within dynamical mean-field theory.
  
\end{abstract}

\maketitle

\section{Introduction}
The numerical solution of the quantum many-body problem out of equilibrium is an
outstanding challenge in modern physics, required to simulate the effect
of strong radiation fields on atoms and
molecules~\cite{balzer2010a,balzer2010b}, quantum
materials~\cite{basov2017towards,giannetti2016ultrafast,ligges2018ultrafast},
nuclear physics~\cite{berges2004,berges2004introduction,bonitz2003proceedings},
ultracold atomic gases~\cite{kinoshita2006,schneider2012,sandholzer2019quantum}, and
many other systems. Various theoretical frameworks for
equilibrium problems have been extended to the nonequilibrium
situation, including density functional
theory~\cite{marques2012fundamentals}, the density matrix
renormalization group (DMRG)~\cite{schollwock2011density}, and field theory
approaches based on the Keldysh
formalism~\cite{kamenev2011field,stefanucci2013nonequilibrium,haug2008quantum,aoki2014}.
A typical limitation is the restriction to rather short propagation times.
This inherent difficulty manifests itself in various forms; for example,
in bond dimension growth in DMRG~\cite{schollwock2011density}, the
dynamical sign problem in Monte Carlo
methods~\cite{werner2010,cohen2015,gull2011}, and memory effects in the
Keldysh
formalism~\cite{kadanoff1962quantum,stefanucci2013nonequilibrium,schuler2020,aoki2014}.
Extending propagation times would allow for the investigation of new
phenomena, such as the stabilization of metastable
states~\cite{fausti2011,mitrano2016,stojchevska2014ultrafast,zhou2019nonequilibrium}, which
take place on time scales that are orders of magnitude larger than those currently reachable by state-of-the-art techniques. 

\begin{figure}[t]
  \centering
  \begin{subfigure}[t]{.23\textwidth}
    \centering
    \includegraphics[width=\linewidth]{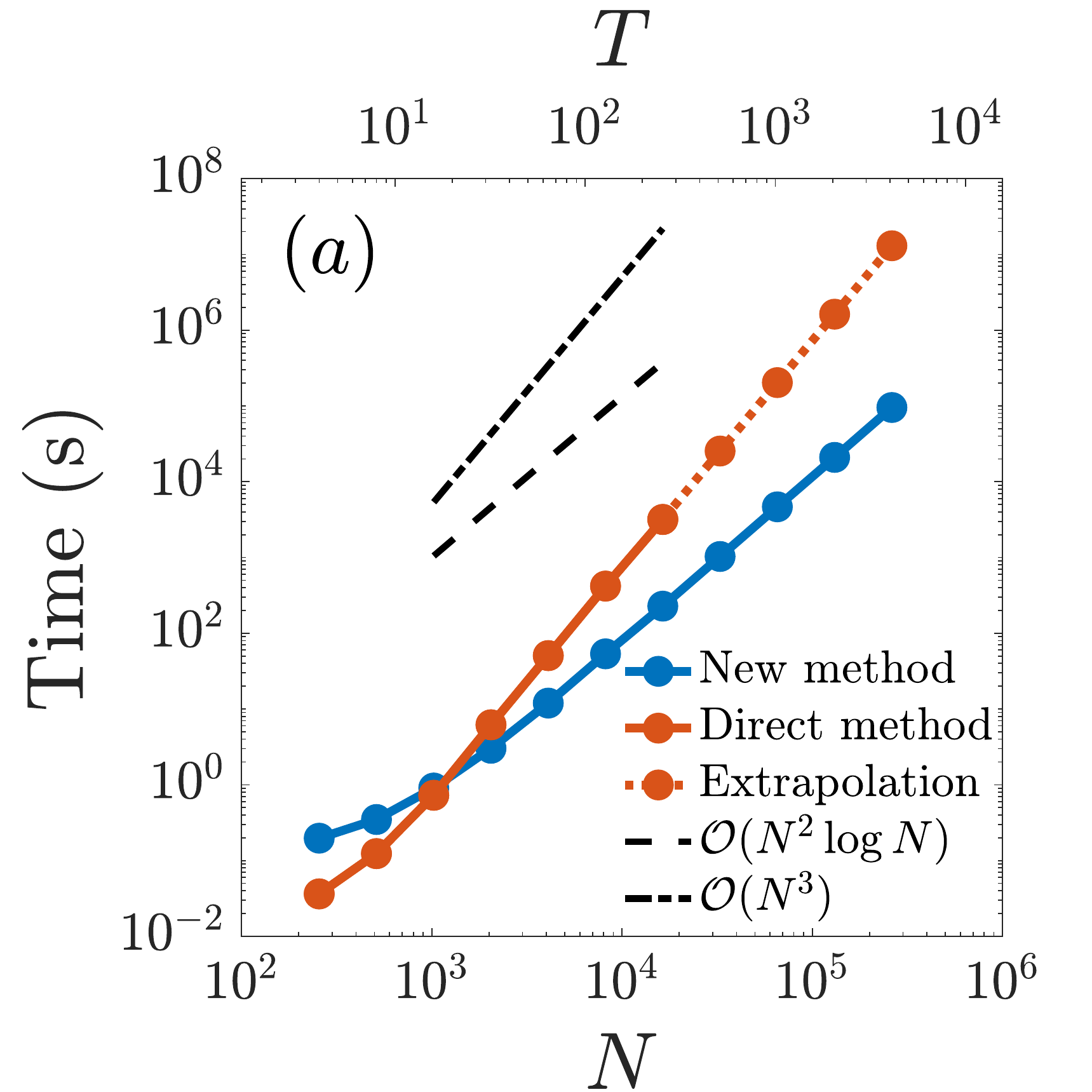}
    \captionlistentry{}
    \label{fig:varyt_time_ramp}
  \end{subfigure}
  \begin{subfigure}[t]{.23\textwidth}
    \centering
    \includegraphics[width=\linewidth]{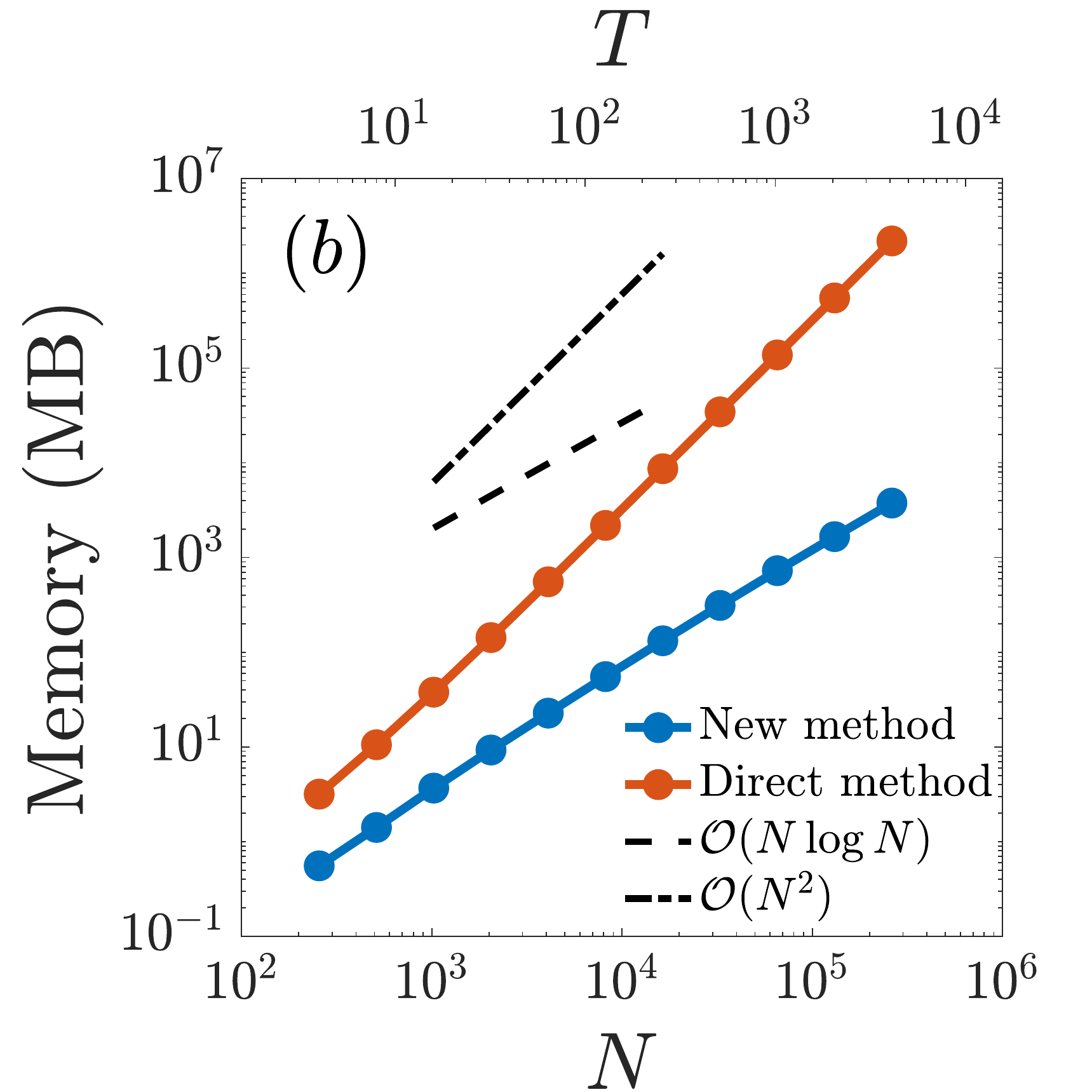}
    \captionlistentry{}
    \label{fig:varyt_mem_ramp}
  \end{subfigure}
  \vspace{-3ex}
  \caption{(a) Wall clock time and (b) memory used by the direct method and our
  method for the
  Falicov-Kimball model with a rapid ramp of the interaction parameter,
  showing the improved scalings.
  The time step $\Delta t$ is fixed, and the number of time steps is
  increased from $N = 256$ to $262\,144$, corresponding to an increase
  of the propagation time from
  $T=4$ to $4096$. The direct method is impractical for large $N$, so we show extrapolated timings.}
\label{fig:varyt_ramp}
\end{figure}

The Keldysh formalism is a particularly versatile approach, as it is not
limited by the dimension of the problem~(like DMRG), and can be
efficiently adjusted to realistic setups. Several recent studies have
used these techniques in direct comparisons with experiments, including
those involving transport properties~\cite{schlunzen2016} and periodic
driving~\cite{sandholzer2019quantum} in ultra-cold atomic systems, and
pump-probe experiments in correlated
solids~\cite{ligges2018ultrafast,gillmeister2020,li2018theory,golez2017,kumar2019higgs,sentef2013}. Many have already
reached the level of first-principles
description~\cite{perfetto2018ultrafast,molina2017ab,sangalli2019many,marini2009yambo,perfetto2018cheers}.
The essential task is to evaluate the two-time Green's function either
with a numerically exact
method~\cite{bertrand2019,macek2020,bertrand2019a} or with a high-order
approximation scheme adjusted for the problem being
studied~\cite{eckstein2010,aoki2014,abdurazakov2018,kemper2018,golevz2019multiband,peronaci2018,peronaci2020,bittner2020effects,bittner2018,golez2017}.

Unfortunately, because of the necessity of computing full history
integrals, the solution of the underlying
Dyson equation by standard algorithms has a computational complexity of
$\OO{N^3}$ and a memory complexity of $\OO{N^2}$ with respect to the
number $N$ of time steps~\cite{stefanucci2013nonequilibrium}. Numerous
proposals have been made to improve efficiency,
including memory
truncation~\cite{schuler2018}, high-order time stepping and quadrature
rules~\cite{schuler2020},
parallelization~\cite{balzer2012nonequilibrium,talarico19}, and direct Monte Carlo
sampling at long times~\cite{bertrand2019,bertrand2019a}. Of these, only
the memory truncation methods succeed in systematically reducing the asymptotic cost and memory
complexity, but these are restricted to specific parameter regimes in
which the Green's functions are numerically sparse
\cite{picano2020accelerated,schuler2018}. 
Another alternative is to approximate the full propagation
scheme with quantum kinetic equations or their generalizations, like
the generalized Kadanoff-Baym
ansatz~(GKBA)~\cite{lipavsky1986,kalvova2019,schlunzen2020}. These
techniques are sufficiently accurate to explore long-time processes in
several setups~\cite{murakami2020ultrafast,schuler2019,schuler2020,karlsson2020}.
However, they rely on an approximation for the equations of motion which
is not always valid~\cite{tuovinen2020comparing,tuovinen2019}. Moreover, they lose their advantageous scaling for higher-order expansions.

There is still a demand, therefore, for a versatile propagation scheme
with reduced computational and memory complexity which is compatible
with recently developed nonperturbative techniques, like time-dependent
dynamical mean-field theory~(DMFT)~\cite{aoki2014,eckstein2010} and
numerically exact Monte-Carlo approaches~\cite{bertrand2019,macek2020,bertrand2019a,werner2010,cohen2015,gull2011}.
In this work, we propose a propagation scheme which, for
systems whose Green's functions and self energies have the so-called
hierarchical off-diagonal low rank (HODLR) property, reduces the computational complexity from
$\OO{N^3}$ to $\OO{N^2 \log N}$ and the memory complexity from
$\OO{N^2}$ to $\OO{N \log N}$. The
HODLR structure allows us to build a compressed representation of Green's
functions and self energies on the fly using the truncated singular
value decomposition, and we use this representation to accelerate
the evaluation of history integrals. We have confirmed the HODLR property in
several systems of physical interest, and present results for the Falicov-Kimball model
and the Hubbard model excited by a rapid ramp or a periodic driving.
Our numerical examples show simulations with unprecendented propagation
times, and computational cost and memory reductions of orders of
magnitude. Scaling results for an example using the Falicov-Kimball
model are shown in Fig. \ref{fig:varyt_ramp}.

Our method may be integrated into existing time stepping schemes,
including high-order discretizations, with user-controllable accuracy.
That is, the additional error is set by a user-defined parameter
$\varepsilon$, and the cost and memory requirements increase slowly as
$\varepsilon$ is decreased for HODLR systems. Efficiency is achieved not
by making additional modeling assumptions, but by exploiting an existing
compressibility structure. Notably, the algorithm
discovers the ranks in the HODLR representation automatically, and if a
system fails to obey the HODLR property to some degree, the algorithm
will simply slow down accordingly in order to guarantee $\varepsilon$
accuracy, rather than give an incorrect result.

This manuscript is organized as follows. In Sec.~\ref{KBE}
we describe the Kadanoff-Baym form of the nonequilibrium Dyson equation
and review the standard method of solving it. In Sec.~\ref{Sec:Fast} we
introduce the HODLR compression structure, show how to build it on the
fly, and describe a fast algorithm for the evaluation of history
integrals. In Sec.~\ref{Examples} we demonstrate a full implementation for the
Falicov-Kimball model, and study the HODLR compressibility of Green's
functions for the
Hubbard model excited by a rapid ramp and by periodic driving of system
parameters. In Sec.~\ref{Conclusion} we summarize our results and
discuss several future directions of research.

\section{The Kadanoff-Baym equations}\label{KBE}
The Keldysh formalism describes the single particle two-time Green's
functions
\begin{align}
  \label{eq:def_green}
  G_{jk}(t,t^\prime) = -i\langle T_\mathcal{C} \hat{c}_j(t)
  \hat{c}^\dagger_k(t^\prime) \rangle \ ,
\end{align}
where $\hat{c}^\dagger_j$ ($\hat{c}_j$) denotes the fermionic or
bosonic creation (annihilation) operator with respect to the $j^\text{th}$
single particle state, and $T_\mathcal{C}$ is the contour order
operator; see Refs.~\onlinecite{stefanucci2013nonequilibrium,schuler2020}. The
construction of the Green's functions is typically carried out by first
evaluating the self energy diagrams defining whichever approximation is
employed, and then solving the Dyson equation, which resums the
subset of diagrams up to infinite order. In this work, we focus on the
second step, and consider situations in which the solution of the Dyson
equation is the computational bottleneck, although our method may still
yield a significant reduction in memory usage in other cases.

The nonequilibrium Dyson equation in the integro-differential form is given by 
\beq{
  \paren{\I \partial_t -h(t)} G(t,t') - \int_{\mathcal{C}} d\bar t \, \Sigma(t,\bar t) G(\bar t,t')= \delta_{\mathcal{C}}(t,t'),
  \label{Eq:Dyson}
}
where $h$ is the single particle Hamiltonian and $\Sigma$ is the
self energy. Here, for simplicity, we consider the scalar case $j=k=1$, but the extension
to the multidimensional case is
straightforward. $\Sigma$ in general depends nonlinearly on $G$. The
Green's function is typically parametrized in terms of Keldysh
components, and for the solution of the Dyson equation, it is
particularly useful to employ a set of physical components: the
Matsubara component $G^M$ , the retarded component $G^R$, the
left-mixing component $G^\rceil$, and the lesser component $G^<$. The
equations of motion for these components lead to the
\textit{Kadanoff-Baym equations}, a set of causal coupled
nonlinear Volterra integro-differential equations (VIDEs) given by~\cite{kadanoff1962quantum,aoki2014,bonitz2003proceedings,stefanucci2013nonequilibrium}
\begin{align}
&\paren{-\partial_{\tau}-h(0)}G^\mat(\tau) -\int_0^\beta d \bar\tau
\,\Sigma^\mat(\tau-\bar\tau)G^\mat(\bar\tau) = 0
\label{eq:dysonm} \\
&\paren{-\I\partial_{t'}-h(t')}G^R(t,t')  -\int_{t'}^t d\bar t \, G^R(t,\bar t)
\Sigma^R(\bar t,t') = 0
\label{eq:dysonr} \\
\begin{split}
  &\paren{\I\partial_t-h(t)}G^{\rceil}(t,\tau)
  - \int_{0}^t d\bar t \,\Sigma^{\mathrm{R}}(t,\bar t ) G^{\rceil}(\bar
  t,\tau) \\ & \hspace{.5cm} = \int_0^{\beta} d \bar\tau \, \Sigma^{\rceil}(t,\bar\tau) G^M(\bar\tau-\tau)
\end{split}
\label{eq:dysonlm} \\
\begin{split}
&\paren{\I\partial_t-h(t)}G^{<}(t,t^\prime)  -
\int_{0}^t d\bar t \,\Sigma^{\mathrm{R}}(t,\bar t ) G^{<}(\bar
 t,t^\prime) \\ &\hspace{.5cm} =
\int_{0}^{t^\prime} d\bar t \,\Sigma^<(t,\bar t ) G^{A}(\bar t,t^\prime) 
-\I \int_0^\beta d \bar\tau \,\Sigma^{\rceil}(t, \bar\tau)
G^{\lceil}_1(\bar\tau,t^\prime)
\end{split}
\label{eq:dysonl}
\end{align}
along with the conditions
\begin{align}
  & G^\mat(-\tau)= \xi G^\mat(\beta-\tau) \label{eq:dysonbcm} \\
  & G^R(t,t) = -i \label{eq:dysonbcr} \\
  & G^{\rceil}(0,\tau)=iG^\mat(-\tau)=i \xi G^\mat(\beta-\tau)
  \label{eq:dysonbclm} \\
  & G^<(0,t^\prime) =-\wb{G^\rceil(t^\prime,0)}
  \label{eq:dysonbcl}
\end{align}
and the relations
\begin{align}
  & G^{\lceil}(\tau,t) = -\xi \wb{G^\rceil(t,\beta-\tau)}
  \label{eq:grmdef} \\
  & G^A(t,t') = \wb{G^R(t',t)}
  \label{eq:gadef} \\
  & G^<(t,t') = -\wb{G^<(t',t)}.
  \label{eq:glantisym}
\end{align}
Here $\xi = \pm 1$ for the bosonic and fermionic cases, respectively,
$\wb{\mbox{$\,\cdot\,$\raisebox{1.5mm}{}}}$ denotes complex
conjugation, $\beta$ is the given inverse temperature, and $\tau$ is an
imaginary time variable. We note that we have used the conjugate equation
for the retarded component in \eqref{eq:dysonr}. For a detailed
derivation of the Kadanoff-Baym equations, we refer the reader to Ref.
\onlinecite{stefanucci2013nonequilibrium}.

\subsection{Direct solution of the Kadanoff-Baym equations}
\label{sec:directmethod}

Our method is built on top of the direct time stepping procedure which
has traditionally been used to solve the Kadanoff-Baym equations. We now
briefly review that procedure. For details about discretization, initialization procedures for high-order methods, nonlinear iteration,
and the evaluation of self energies, we refer the reader to Refs.~\onlinecite{stefanucci2013nonequilibrium,schuler2020}.

We assume a discretization of the variables $t$ and $t'$ on a uniform
grid $t_n = n \Delta t$ with $n = 0,1,\ldots,N$, and of the variable
$\tau$ on a uniform grid $\tau_k =
k \Delta \tau$ with $m = 0,1,\ldots,M$. The final time is given by $T = N \Delta t$, and the inverse
temperature by $\beta = M \Delta \tau$. The Green's
functions are sampled on the appropriate products of these grids to
form matrices, except for the Matsubara component, which is only a
function of $\tau$. The retarded Green's function is represented by a
lower triangular matrix, and because of its Hermitian antisymmetry
\eqref{eq:glantisym}, the
lesser Green's function is determined by its lower or upper triangular
part~\cite{schuler2020}. 

First, Eqns. \eqref{eq:dysonm} and \eqref{eq:dysonbcm} for the Matsubara component may be
solved independently of the other components. Several efficient
numerical methods exist \cite{dong2020,gull2018chebyshev,schuler2020},
and we do not consider this topic here.

The entries for each of the other Green's functions are computed in the following
order:
\begin{itemize}
  \item The lower triangular matrix
    $\GR(t_m,t_n)$ is filled in with $m$
    proceeding from $0$ to $N$ in an outer iteration, and with $n$
    proceeding from $m$ to $0$ in an inner iteration.
  \item The rectangular matrix
    $\GLM(t_m,\tau_k)$ is filled in with $m$ proceeding from $0$ to $N$
    and for each $k$ in parallel.
  \item The upper-triangular matrix
    $\GL(t_n,t_m)$ is filled in with $m$
    proceeding from $0$ to $N$ in an outer iteration, and with $n$
    proceeding from $0$ to $m$ in an inner iteration.
\end{itemize}
More specifically, suppose we have reached the outer time step $t_{m'} =
m' \Delta t$, so that $\GR(t_m,t_n)$, $\GLM(t_m,\tau_k)$, and $\GL(t_n,t_m)$ are
known for $m = 0,1,\ldots, m'-1$, $n = 0,1,\ldots,m$, and $k =
0,1,\ldots,M$. We can then fill in the matrix entries corresponding to
$m = m'$, $n = 0,1,\ldots,m'$, and $k = 0,1,\ldots,M$. Since the
self energies depend in general on the values of the Green's functions at
these points, we must carry out a self-consistent iteration on the new
entries of the Green's functions. At the beginning of each iterate, the new entries
of the self energies are first computed based on some combination of extrapolation
from previous time steps and previous iterates of the current time step.
Then, assuming fixed self energies, the new entries of the Green's
functions for a given iterate are computed by the following procedure:

\begin{enumerate}
  \item The equation \eqref{eq:dysonr} for the retarded component is a linear VIDE in
    $t'$ for fixed $t = t_{m'}$. It is solved by a time stepping
    procedure, starting at $t' = 
    t_{m'} = t$, where we use the condition \eqref{eq:dysonbcr}, backwards to $t'
    = t_0 = 0$. The result is the new row
    $\{\GR(t_{m'},t_n)\}_{n=0}^{m'}.$ 
  \item The equation \eqref{eq:dysonlm} for the left-mixing component is
    a system of coupled
    linear VIDEs indexed by $\tau = \tau_k$. It is solved for
    the new row $\{\GLM(t_{m'},\tau_k)\}_{k=0}^M$. Note that at the
    first time step, we use the initial condition \eqref{eq:dysonbclm}.
  \item The equation \eqref{eq:dysonl} for the lesser component is a linear VIDE in $t$ for fixed
    $t' = t_{m'}$. It is solved starting at $t = t_0 = 0$, where we have the
    now known condition \eqref{eq:dysonbcl}, forwards to $t =
    t_{m'} = t'$. The result is the new column
    $\{\GL(t_n,t_{m'})\}_{n=0}^{m'}$. Note that as a consequence of the
    Hermitian antisymmetry \eqref{eq:glantisym} of $G^<$ and the
    definitions \eqref{eq:grmdef} and \eqref{eq:gadef}, the right hand
    side of \eqref{eq:dysonl} is entirely known.
\end{enumerate}

The most expensive step in the solution of each VIDE is the evaluation
of the various integrals at each time step. We refer to these
as history integrals, or history sums when discretized,
since they involve summation over
previously computed quantities. To understand the cost, we first
consider the history integral in the VIDE for the retarded
component corresponding to the outer time step $t = t_m$ and inner time
step $t = t_n$, discretized by the trapezoidal rule as 
\begin{equation} \label{eq:histret}
  I_{m,n}^{R,1} = \Delta t \sump{j=n}{m} G^R(t_m,t_j) \Sigma^R(t_j,t_n) .
\end{equation}
Here, the primed sum symbol indicates that the first and last terms of
the sum are weighted by $1/2$. The specific discretization used is
unimportant for our discussion, and we have chosen the trapezoidal rule
for simplicity. For each $m = 0,\ldots,N$, we must compute this sum for
$n = 0,\ldots,m$, so that the cost of computing all such sums scales as
$\OO{N^3}$. By contrast, the cost of time stepping for all outer time
steps $t_m$ and inner time steps $t_n$, ignoring the history sums,
scales as $\OO{N^2}$. Furthermore, computing these sums requires storing
$\Sigma^R(t_m,t_n)$ in its entirety, an $\OO{N^2}$ memory cost.

In Table \ref{tab:histsums}, we list the six history sums, obtained by discretizing the corresponding
integrals in Eqns. \eqref{eq:dysonr}--\eqref{eq:dysonl}, along with the
total cost of computing them directly. Each history sum is slightly different, but for each
Keldysh component, the cost of computing the history sums is dominant.
Furthermore, in order to compute the sums, one must store all of the
computed Green's functions and/or the corresponding self energies. Our
main objective is to reduce these bottlenecks.

\begin{table}[t]
  \centering
  \footnotesize
  \begin{tabular}{|c|M{1.5cm}|c|}
    \hline
    History sum & Direct summation cost & Fast summation cost \\ \hline
    $I_{m,n}^{R,1} = \Delta t \sumplim{j=n}{m} G^R(t_m,t_j) \Sigma^R(t_j,t_n)$ &
    $\OO{N^3}$ & $\OO{k N^2 \log N + k^2 N^2}$ \\ \hline

    $I_{m,k}^{\rceil,1} = \Delta t \sumplim{j=0}{m} \Sigma^R(t_m,t_j)
    \GLM(t_j,\tau_k) $ & 
    $\OO{N^2 M}$ & $\OO{kN^2+kNM+k^2N}$ \\ \hline

    $I_{m,k}^{\rceil,2} = \Delta \tau \sumplim{l=0}{M}
    \Sigma^\rceil(t_m,\tau_l) G^M(\tau_l-\tau_k) $ &
    $\OO{N M^2}$ & $\OO{NM \log M}$ \\ \hline
    
    $I_{n,m}^{<,1} = \Delta t \sumplim{j=0}{n} \Sigma^R(t_n,t_j) \GL(t_j,t_m)$ &
    $\OO{N^3}$ & $\OO{k N^2 \log N + k^2 N^2}$ \\ \hline
    
    $I_{n,m}^{<,2} = \Delta t \sumplim{j=0}{m} \Sigma^<(t_n,t_j) G^A(t_j,t_m)$ &
    $\OO{N^3}$ & $\OO{k N^2 \log N + k^2 N^2}$ \\ \hline
    
    $I_{n,m}^{<,3} = \Delta \tau \sumplim{l=0}{M} \Sigma^\rceil(t_n,\tau_l)
    G^\lceil(\tau_l,t_m)$ & $\OO{N^2 M}$ & $\OO{kN^2 + kNM + k^2N}$ \\

    \hline
  \end{tabular}
  \caption{History sums and the total cost of computing them for
  all index values by direct summation and by our method. The sums are trapezoidal rule discretizations of
  the underlying integrals. $k$ is a
  bound on the ranks of all blocks in the compressed representations. The
  total cost of building these representations by on the fly TSVD
  updates is $\OO{k^2 \paren{N^2 + NM}}$.}
  \label{tab:histsums}
\end{table}

Before discussing our approach, it will be useful to understand the history sums in
terms of matrix algebra. We again use the retarded component as an
example. At each outer time step $t_m$,
the collection of sums $\{I_{m,n}^{R,1}\}_{n=0}^m$ may be viewed as the
product of a $1 \times m$ row vector $\{\GR(t_m,t_j)\}_{j=0}^m$ with an $m \times m$
lower triangular matrix ${\{\Sigma^R(t_j,t_n)\}_{j=0}^{m}} \,
{\vphantom{\{\Sigma^R(t_j,t_n)\}}_{n=0}^{j}}$, properly modified to
take the trapezoidal rule weights into account. The cost of computing
each such product is $\OO{m^2}$, so that the cost of computing all such
products is $\OO{N^3}$. Of course, we cannot
compute all sums simultaneously by such a matrix-vector product, since
$\{\GR(t_m,t_j)\}_{j=0}^m$ is itself built during the course of time stepping.
Rather, this product is computed one step at a time; at the time step $t_n$, we compute the product of the row vector
with the $n^\text{th}$ column of the lower triangular matrix.

\section{A fast, memory-efficient method based on hierarchical low rank compression}\label{Sec:Fast}

To reduce the bottlenecks associated with history storage and summation,
we might first hope that the Green's functions and self energies display
some sort of sparsity. For example, if the retarded Green's function decays rapidly
in the off-diagonal direction, we do not need to store its entries with
moduli smaller than some threshold, and we can ignore them in the history
sums. Though this is sometimes the case~\cite{schuler2018,picano2020accelerated}, decay in the Green's
functions and self energies depends strongly on the parameter regime. However,
we have found that
the Green's functions and self energies for many systems of physical
interest display a smoothness property which leads to
\textit{data sparsity}; they have numerically low rank off-diagonal
blocks. This can be systematically exploited to achieve highly compressed
representations which admit simple algorithms for fast matrix-vector
multiplication.

We will first discuss the compressed storage format for the Green's
functions and self energies. Then we will describe how to build these
compressed representations on the fly, as new matrix entries are filled
in. Finally, we will show that the compressed format leads naturally to
a fast algorithm to compute the history sums.
\begin{figure}[t]
  \centering
    \includegraphics[width=0.75\linewidth]{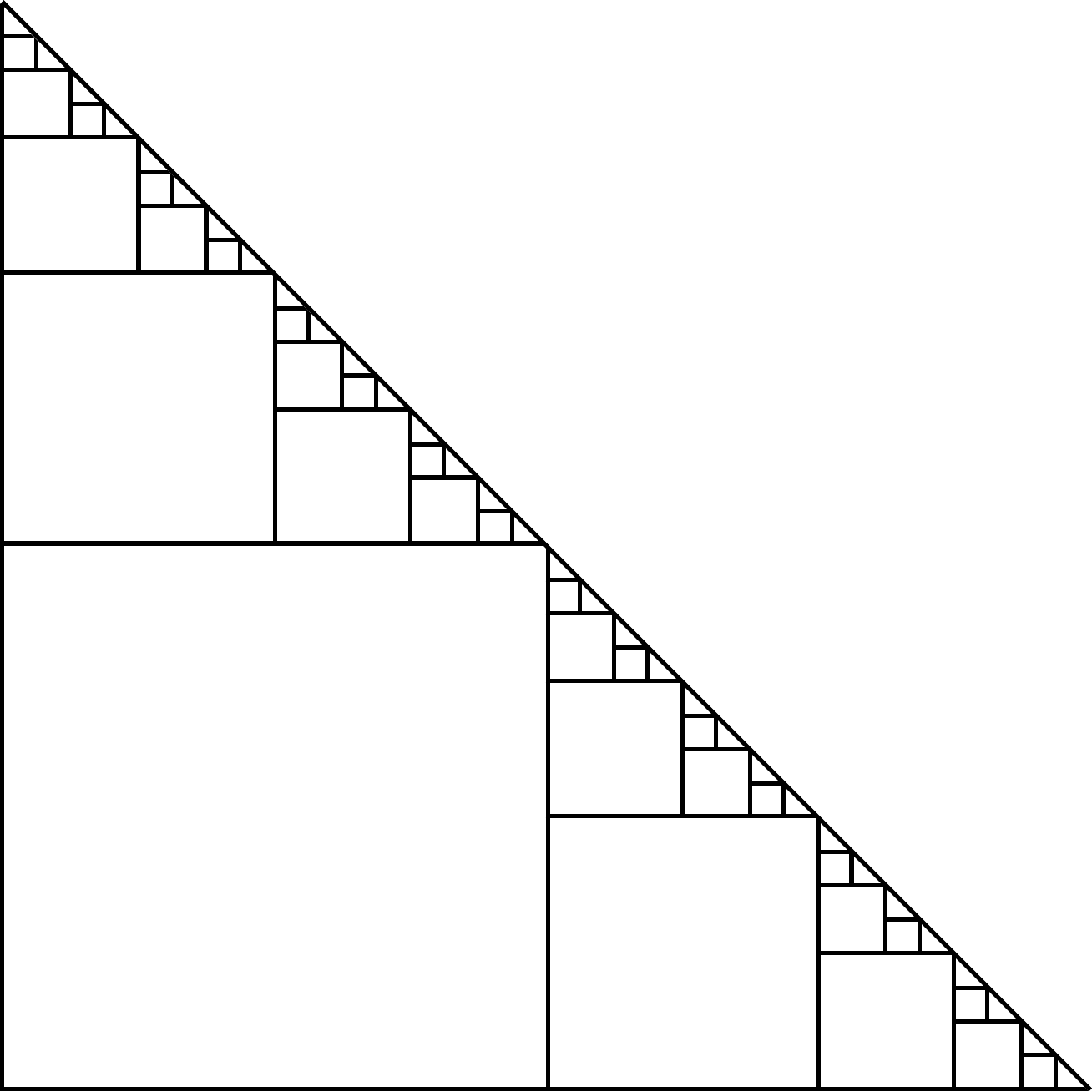}
    \caption{Hierarchical off-diagonal partioning of a lower triangular
    $N \times N$ matrix. The matrix is HODLR if the blocks in the recursive partitioning above, continued towards the diagonal until the smallest blocks are of a small constant size, have
    $\varepsilon$-rank $k \ll N$.}
    \label{fig:triblock}
\end{figure}
\subsection{Hierarchical low rank compression of Green's functions}\label{Sec:HODLR}

Consider first the retarded Green's function, which is discretized as a
lower triangular matrix. We partition the matrix into blocks by
recursive subdivision, as in
Figure \ref{fig:triblock}. Each
block may be described by its \textit{level}; we
say that the largest block is in level one, the two second
largest blocks are in level two, and so on, with $L$ levels
in total. We choose $L \sim \log_2 N$ so that each of the triangular blocks near the
diagonal contains a small, constant number of entries.
All blocks are
approximately square, and the blocks in a given level
have dimensions approximately half of those in the previous level.

A matrix is said to have the \textit{hierarchical off-diagonal low rank (HODLR)
property with $\varepsilon$-rank $k$} if each of the blocks in this
partitioning are rank $k$ to a threshold $\varepsilon$ -- that is, their
$(k+1)^\text{th}$ singular values are less than $\varepsilon$ \cite{ballani16}.
Using the truncated singular value decomposition (TSVD), obtained from
the SVD by setting all singular values less than $\varepsilon$ to zero
and deleting the corresponding left and right singular vectors, an $n
\times n$ block with $\varepsilon$-rank $k$ may be stored using 
$k (2n+1)$ rather than $n^2$ numbers, and recovered with error at most
$\varepsilon$ in the spectral norm. The TSVD is the best rank $k$
approximation in this norm, with error given by the $(k+1)^\text{th}$ singular value
\cite[Sec. 2, Thm. 2]{ballani16}. Since each of the
$2^{l-1}$ blocks in
level $l$ has dimensions $n \approx N/2^l$, all blocks in a HODLR matrix may be stored
with arbitrary accuracy $\varepsilon$ using approximately
\[k \sum_{l=1}^L 2^{l-1} (N/2^{l-1}+1) = \OO{k N
\log_2 N} \]
rather than $\OO{N^2}$ numbers. The entries in the triangular blocks
near the diagonal may be stored directly, and since we choose $L \sim
\log_2 N$, there are only $\OO{N}$ of them.

We are primarily interested in the behavior of the family of matrices obtained by
fixing $\Delta t$
and increasing $N$ to reach longer propagation times. It may be
that in this family, the $\varepsilon$-rank bound $k$ itself increases with $N$. If $k$ grows
linearly with $N$, then there is no asymptotic advantage to storing the
matrix in the compressed format described above. If $k$ does not grow with $N$ at all -- the ideal
case -- then the
total cost of storage in the compressed format grows only as $\OO{N
\log N}$. We have examined the $\varepsilon$-rank behavior of retarded Green's
functions and self energies in the compressed format for a variety of
physical systems. \textit{Our crucial empirical observations are, first,
that for fixed $N$, the maximum $\varepsilon$-rank $k$ of any block is often much less
than $N$, and second, that in these cases $k$ grows only weakly with $N$, so that
storage costs are close to $\OO{N \log N}$ with a small scaling constant.}

The matrices of the lesser Green's function and self energy are
Hermitian antisymmetric, so we need only store their lower triangular parts. We
have observed that they have similar $\varepsilon$-rank structures to the retarded components. The left-mixing Green's function
is represented by a full $N \times M$ matrix, and our observation is
that this matrix is often simply of low $\varepsilon$-rank. Using the TSVD, then, we can
store it using $k (N + M + 1)$ rather than $NM$ numbers, where here and
going forward we use $k$ to denote an $\varepsilon$-rank bound for all
Keldysh components.

We note that if $T$ is fixed and $N$ is increased in order to
achieve higher resolution -- the $\Delta t \to 0$ regime -- the $\varepsilon$-ranks cannot
increase asymptotically with $N$, and we are guaranteed $\OO{N \log N}$
scaling of the memory usage.
In this case, the size of the constant $k$ in the scaling
entirely determines the efficiency of the method.

The rank properties, and consequently the compressibility, of the
Green's functions and self energies vary from system to system.
We give numerical evidence of significant compressibility for several
systems of physical interest in Section~\ref{Examples}.

\subsection{Building the compressed representation on the fly}
\label{sec:tsvdupdate}

If we are required to construct each block in its entirety before
compressing it, the peak memory usage will still scale as $\OO{N^2}$.
Therefore, we must build the Green's functions and self energies in the compressed format
on the fly. During the course of time stepping, new rows of the
retarded Green's function matrix are filled in one at a time, according
to the procedure described in
Sec. \ref{sec:directmethod}. Each new row may be
divided among the blocks in the HODLR partition containing a part of it. The new entries
in the triangular blocks are stored directly. For the
other blocks, we must compute the TSVD of the concatenation
of a block known in TSVD representation with a new row. We carry out
this process, called an SVD update, using the method described in Ref.~\cite[Sec. 2-3]{brand06}, which we outline below.

Suppose we have a matrix $B'$ given by the first $m'$ rows of
an $m \times n$ block $B$, with rank $k'$ TSVD $B' = U' S' V'^*$.
Given a new row $b^*$ of $B$, we wish to construct the
$\varepsilon$-rank TSVD of the
matrix
\[B^+ = 
  \paren{
    \begin{array}{c}
      B' \\
      \hline
      b^*
    \end{array}
  }
\]
in an efficient manner -- in particular, we cannot simply expand $B'$
from its TSVD, append $b^*$, and compute the TSVD of $B^+$ directly,
since this would lead to $\OO{N^2}$ peak memory usage. If $m' = 0$, $B'$
is empty, and the TSVD is simply given by $B^+ = U^+ S^+ V^{+*}$ with
$U^+ = 1$, $S^+ = 1$, and $V^+ = b$. Otherwise, we begin by writing 
\[B^+ = 
  \paren{
    \begin{array}{c|c}
      U' & 0 \\
      \hline
      0 & 1
    \end{array}
  }
  \paren{
    \begin{array}{c|c}
      S' & 0 \\
      \hline
      0 & 1
    \end{array}
  }
  \paren{
    \begin{array}{c}
      V'^* \\
      \hline
      b^*
    \end{array}
  }.
\]
We first orthogonalize the new row against the current
row space of $B$, which is given by the span of the columns of $V'$. In
particular, define the normalized orthogonal complement $q = (b -
V'V'^*b)/\beta$ with $\beta = \norm{b -
V'V'^*b}.$ $V'^*b$ and $q$ may be computed at a cost of $\OO{n k'}$
using the modified Gram-Schmidt algorithm~\cite[Sec. 5.2.8]{golub96}. We then have
\[
  \paren{
    \begin{array}{c|c} V' & b \end{array}
    }
  =
  \paren{
    \begin{array}{c|c} V' & q \end{array}
    }
  \paren{
    \begin{array}{c|c}
      I & V'^* b \\
      \hline
      0 & \beta
    \end{array}
  }
\]
and therefore
\begin{equation} \label{eq:bplusnew}
B^+ = 
  \paren{
    \begin{array}{c|c}
      U' & 0 \\
      \hline
      0 & 1
    \end{array}
  }
  \paren{
    \begin{array}{c|c}
      S' & 0 \\
      \hline
      b^* V' & \beta
    \end{array}
  }
  \paren{
    \begin{array}{c}
      V'^* \\
      \hline
      q^*
    \end{array}
  }.
\end{equation}
The middle matrix is a $k'+1 \times k'+1$
half-arrowhead matrix, and we can compute its SVD $U S V^*$ in
$\OO{k'^2}$ time \cite{stor15}. The rank $k'+1$ TSVD of $B^+$ is then given by $B^+ = U^+ S^+
  V^{+*}$, with $U^+ = \paren{ \begin{array}{c|c} U' & 0 \\ \hline 0 & 1
\end{array}} U$, $S^+ = S$, and $V^+ = \paren{ \begin{array}{c|c} V' & q
  \end{array}} V$. The cost of 
  forming $U^+$ and $V^+$ by matrix-matrix multiplication is
  $\OO{k'^2 (m'+n)}$, and is the asymptotically dominant cost in the update. If $S^+_{k'+1,k'+1} < \varepsilon$, then
$B^+$ has $\varepsilon$-rank $k'$, and we
remove the last column of $U^+$, the last row of $V^+$, and the last
row and column of $S^+$.

The cost of building a full $m \times n$ block of rank at most $k$ one
row at a time in this
manner is $\OO{k^2 m (m+n)}$. For the retarded and lesser Green's functions and self energies, in each block we have $m \approx n$, with $2^{l-1}$ blocks at
level $l$ of dimensions $n \approx N/2^l$, so the total update cost is of the
order
\[k^2 \sum_{l=1}^L 2^{l-1} \paren{N^2/2^{2l}} = \OO{k^2 N^2}.\]
The left-mixing Green's function and self energy are 
$N \times M$ matrices with rows filled in one by one, so we can use the
update procedure for a single block, at a total cost of $\OO{k^2 \paren{N^2+NM}}$.

\subsection{Fast evaluation of history sums for the retarded
component} \label{sec:fasthistret}

\begin{figure}[t]
  \centering
    \includegraphics[width=0.8\linewidth]{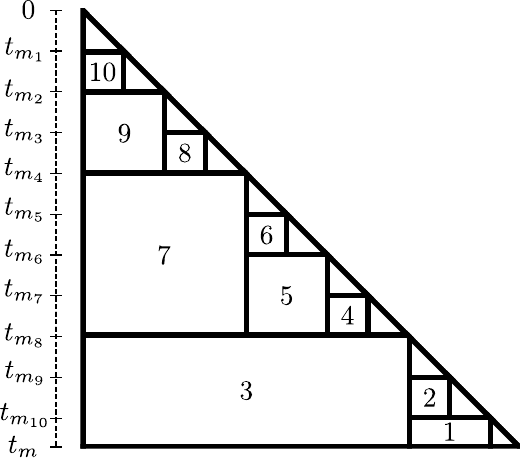}
    \caption{The portion of compressed representation 
    depicted in Figure \ref{fig:triblock} which has been constructed by
    the outer time step $t = t_m$ for the retarded Green's function and
    self energy.
    The block numbering indicates the order in which blocks are applied
    in order to compute the history sums $I_{n,m}^{R,1}$.}
    \label{fig:truncblock}
\end{figure}

Fig. \ref{fig:truncblock} depicts the portion of the compressed representation
of the retarded Green's function and self energy that has been built by the time we have
reached the outer time step $t = t_m$. For the self energy, this is the
compressed representation of the integral
kernel appearing in the VIDE \eqref{eq:dysonr} corresponding to $t = t_m$. We describe
in this section how to use this compressed representation to efficiently compute the
history sums $I_{n,m}^{R,1}$ defined in \eqref{eq:histret}, for $n = 0,\ldots,m$.

On the left side of Fig. \ref{fig:truncblock}, we show a division of
the interval $[0,t_m]$ into panels with endpoints $0, t_{m_1}, t_{m_2},
\ldots, t_{m_{10}}, t_m$. $m_1, m_2, \ldots, m_{10}$ are the
first row indices of the $10$ blocks, ordered from top to bottom. As
described in Section \ref{sec:directmethod}, when solving the 
VIDE corresponding to $t = t_m$ by time stepping, the row vector
$\{\GR(t_m,t_j)\}_{j=0}^m$ is filled in from right to left; we start with
$\GR(t_m,t_m) = -i$, then we fill in $\GR(t_m,t_{m-1})$, and so on,
until we reach $\GR(t_m,0)$. As shown in Eq.~\eqref{eq:histret}, the history sum
$I_{m,n}^{R,1}$ corresponding to time step $t' = t_n$ is given by the
weighted product of the row vector $\{\GR(t_m,t_j)\}_{j=n}^m$ with the column
vector $\{\Sigma^R(t_j,t_n)\}_{j=n}^m$.

From time steps $t_m$ until $t_{m_{10}}$, the portion of the kernel matrix
contributing to the history sums is the lower right triangular block, and we
compute the sums directly. However, once we fill in
$\GR(t_m,t_{m_{10}})$, we can apply the block labeled by $1$ to the row
vector $\{\GR(t_m,t_j)\}_{j={m_{10}}}^m$, yielding a partial contribution
to the history sums $I_{m,n}^{R,1}$ with $n$ indexing the columns of
that block. These contributions are stored. Once we have filled in $\GR(t_m,t_{m_9})$, we can apply the block
labeled by $2$ to the row vector $\{\GR(t_m,t_j)\}_{j={m_9}}^{m_{10}}$,
yielding partial contributions to the history sums corresponding to its
columns, which are added to the previous contributions and stored. Once
we have filled in $\GR(t_m,t_{m_8})$, we can apply the block labeled by
$3$ to the row vector $\{\GR(t_m,t_j)\}_{j={m_8}}^{m}$, again obtaining part
of the corresponding history sums. We
proceed in this manner, applying blocks as soon as all
entries of $\{\GR(t_m,t_j)\}_{j=0}^{m}$ corresponding to the block row indices
become available, and adding the result to the history sums
corresponding to the block column indices.

By the time we reach some time step $t_n$, we will have already computed
$I_{m,n}^{R,1}$, except for the \textit{local part} of the sum; that is,
the product of the part of the $n$th column contained in a triangular
block with the corresponding part of $\{\GR(t_m,t_j)\}_{j=0}^m$, which
contains the most recently added entries. We compute this small dot
product directly and add it in to obtain the full history sum $I_{m,n}^{R,1}$.

So far, we have not described a fast algorithm, only a way of
reorganizing the computation of
the history sums into a collection of matrix-vector products and small
dot products. To obtain a fast algorithm, we recall
that each block $B$ is stored as a TSVD $B = U S V^*$, with the
$\varepsilon$-rank bound $k$. If $B$ is $m
\times n$, then it can be applied with a cost scaling as $\OO{k (m+n+k)}$ rather than
$\OO{mn}$ by applying the factors of the TSVD one at a time. Suppose
then that the blocks, including partial blocks, in the compressed
representation of the kernel at the current outer time step $t = t_m$
are in levels at least $l'$. Since there are $\sim 2^l m/N$ blocks each of
dimensions $N/2^l$ at
level $l$, the cost of applying all blocks at that time
step is bounded asymptotically by
\[k \sum_{l'}^{L} \frac{m}{N} 2^l \paren{N/2^l + k} = k m(L-l'+1) +
  \frac{2mk^2}{N} \paren{2^{L+1}-2^{l'}}.\]
Since there are approximately $N/2^{l'-1}$ time steps $t_m$ for which the blocks
are in levels at least $l'$, and $m \leq N/2^{l'-1}$ for such a time step, the total cost of applying all blocks at
every time step is therefore
\begin{multline*}
  k \sum_{l'=1}^L \frac{N}{2^{l'-1}} \paren{ \frac{N}{2^{l'-1}}(L-l'+1) +
  \frac{2k}{N} \frac{N}{2^{l'-1}} \paren{2^{L+1}-2^{l'}}} \\ 
  = \OO{k N^2 \log_2 N + k^2 N^2}.
\end{multline*}
The total cost of computing the local sums is only $\OO{N^2}$.

\subsection{Fast evaluation of history sums for the other
components} \label{sec:fasthistrest}

Each of the history sums defined in Table \ref{tab:histsums} is evaluated by a slightly different algorithm,
but using similar ideas.

\subsubsection{The sum \texorpdfstring{$I_{m,k}^{\rceil,1}$}{Irceil2}}
At the outer time step $t_m$, we compute the product of the row
vector $\{\Sigma^R(t_m,t_j)\}_{j=0}^m$ with the rectangular matrix
${\{\GLM(t_j,\tau_k)\}_{j=0}^{m}} \,
{\vphantom{\{\GLM(t_j,\tau_k)\}}_{k=0}^{M}}$, which is stored as a TSVD,
to obtain $I_{m,k}^{\rceil,1}$ for $k = 0,\ldots,M$.
Applying the factors of the TSVD one by one, we can carry out this
product in $\OO{k (m+M+k)}$ operations, which gives a total
cost for all sums scaling as $\OO{kN^2+kNM+k^2N}$.

\subsubsection{The sum \texorpdfstring{$I_{m,k}^{\rceil,2}$}{Irceil1}}

We compute the product of the row
vector $\{\Sigma^\rceil(t_m,\tau_l)\}_{l=0}^M$ with the square matrix
${\{G^M(\tau_l-\tau_k)\}_{l=0}^{M}} \,
{\vphantom{\{G^M(\tau_l-\tau_k)\}}_{k=0}^{M}}$ to obtain
$I_{m,k}^{\rceil,2}$ for $k = 0,\ldots,M$.
Here, we make use of a different sort of fast algorithm. ${\{G^M(\tau_l-\tau_k)\}_{l=0}^{M}} \,
{\vphantom{\{G^M(\tau_l-\tau_k)\}}_{k=0}^{M}}$ is an $M \times M$ Toeplitz matrix, and
therefore can be applied in $\OO{M \log M}$ time using the fast Fourier
transform (FFT) \cite[Sec 4.7.7]{golub96}.
Briefly, this algorithm works by embedding the Toeplitz matrix in a
larger circulant matrix, zero-padding the input vector, conjugating
the circulant matrix by the discrete Fourier transform (DFT), which
diagonalizes it, and applying the DFT and its inverse using the FFT.
Using this algorithm gives a total cost of $\OO{N M \log M}$ for all the sums.

\subsubsection{The sum \texorpdfstring{$I_{n,m}^{<,1}$}{histles1}}\label{sec:histles1}

The sums $I_{n,m}^{<,1}$ for $n = 0,1,\ldots,m$ are given by the product of the lower
triangular matrix ${\{\Sigma^R(t_n,t_j)\}_{n=0}^{m}} \,
{\vphantom{\{\Sigma^R(t_n,t_j)\}}_{j=0}^{n}}$, stored in the compressed
representation, with the column vector
$\{G^<(t_j,t_m)\}_{j=0}^n$. As for the retarded case, this column vector
is filled in one entry at a time during the course of time stepping,
from $j = 0$ to $j = m$. The algorithm to compute these sums on the fly
is then analogous to that for the retarded case, except that the blocks
are applied in order of increasing column index rather than decreasing
row index as depicted in Figure \ref{fig:truncblock}. The asymptotic
cost is the same as for $I_{n,m}^{R,1}$.

\subsubsection{The sum \texorpdfstring{$I_{n,m}^{<,2}$}{Iles2}}

We compute the product of the square
matrix ${\{\Sigma^<(t_n,t_j)\}_{n=0}^{m}} \,
{\vphantom{\{\Sigma^<(t_n,t_j)\}}_{j=0}^{m}}$ with the column vector
$\{G^A(t_j,t_m)\}_{j=0}^m$ to obtain $I_{n,m}^{<,2}$ for
$n = 0,\ldots,m$. We have 
$\{G^A(t_j,t_m)\}_{j=0}^m = \{\wb{G^R(t_m,t_j)}\}_{j=0}^m $, which is known, and ${\{\Sigma^<(t_n,t_j)\}_{n=0}^{m}} \,
{\vphantom{\{\Sigma^<(t_n,t_j)\}}_{j=0}^{m}}$ is Hermitian
antisymmetric, and therefore also fully
known in the compressed representation. The same
procedure as in Section \ref{sec:histles1} can be used to apply each block of the
lower triangular part, and the
upper triangular part can be applied simultaneously using the
anti-conjugate transposes of the TSVDs of each block. The total asymptotic cost is
therefore again the same as for $I_{n,m}^{R,1}$.

\subsubsection{The sum \texorpdfstring{$I_{n,m}^{<,3}$}{Iles3}}

We compute the product of the matrix ${\{\Sigma^\rceil(t_n,\tau_l)\}_{n=0}^{m}} \,
{\vphantom{\{\Sigma^\rceil(t_n,\tau_l)\}}_{l=0}^{M}}$ with 
the column vector $\{G^\lceil(\tau_l,t_m)\}_{l=0}^M$ to obtain $I_{n,m}^{<,3}$ for $n = 0,\ldots,m$. We perform the matrix-vector product using the TSVD of
${\{\Sigma^\rceil(t_n,\tau_l)\}_{n=0}^{m}} \,
{\vphantom{\{\Sigma^\rceil(t_n,\tau_l)\}}_{l=0}^{M}}$, at a cost of
$\OO{k(m+M+k)}$, giving a total cost of $\OO{kN^2+kNM+k^2N}$ for all
sums.

\subsection{Summary of the time stepping algorithm}

The tools we have described can be integrated into the direct solution
method discussed in Section \ref{sec:directmethod} with only a couple of
modifications:
\begin{itemize}
  \item At the end of self-consistent iteration for each outer time
    step, the TSVD update algorithm described in Section
    \ref{sec:tsvdupdate} must be used to add the new rows of all
    Green's functions and self energies to their compressed
    representations.

  \item The fast procedures described in
    Sections \ref{sec:fasthistret} and \ref{sec:fasthistrest} must be
    used to compute all history sums.
\end{itemize}

By the end of the procedure, we will have computed compressed
representations of each of the Green's functions and self energies.
Operations may be carried out by working directly with the compressed
representations, as in Secs. \ref{sec:fasthistret} and
\ref{sec:fasthistrest}. An entry of a matrix stored in compressed format may be recovered in
$\OO{k}$ operations.

The costs associated with computing the history sums and updating the
compressed representations are summarized in Table \ref{tab:histsums} and its
caption. The storage costs scale as $\OO{k \paren{N \log N + M}}$.

\section{Numerical results}\label{Examples}
In this section, we demonstrate a full implementation of our method for a driven Falicov-Kimball model in the DMFT limit, using two nonequilibrium protocols: a fast ramp and periodic driving. We also test the efficiency of HODLR 
compression offline for the weak and strong coupling regimes of the
Hubbard model. The weak coupling regime is described within the
time-dependent GW approximation for a one dimensional system. The strong
coupling~(Mott insulating) regime is described within the DMFT
approximation on the Bethe lattice with a non-crossing
approximation~(NCA) as the impurity solver.

\subsection{Full implementation for the Falicov-Kimball model}
The Falicov-Kimball problem~\cite{falicov1969,ramirez1970metal}
describes a lattice composed of itinerant $c$ electrons and immobile $f$
electrons which interact via a repulsive Coulomb potential with strength U. The Hamiltonian is given by
\beq{
  H=- J \sum_{\langle i,j\rangle} c_{i}^{\dagger} c_j+ \epsilon \sum_i f_i^{\dagger} f_i + U \sum_{i} f_i^{\dagger} f_i c_i^{\dagger} c_i,
}
where we measure energies in the units of the hopping parameter $J$ and $\epsilon$ is the on-site energy. In the DMFT limit, the equilibrium phase diagram includes metallic, insulating, and charge density wave phases~\cite{freericks2003}. The effective local action for the itinerant electrons is quadratic and the
problem can be solved numerically
exactly~\cite{brandt1989,van1990exact}, so the main computational
bottleneck is the solution of the Dyson equation~\cite{eckstein2008,freericks2008}.

For simplicity, we will
assume the Bethe lattice at half-filling, for which we obtain a pair of coupled equations of the form
\eqref{eq:dysonm}--\eqref{eq:glantisym} for
Green's functions $G_1$ and $G_2$, with 
  \[h_1(t) = U(t)/2, \quad h_2(t) = -U(t)/2,\] 
and $\Sigma$ replaced by
\[\Delta(t,t') = \paren{G_1(t,t') + G_2 (t,t')}/2\]
for each Keldysh component of $G$. $\Delta$ is the hybridization function which enters the solution of the Kadanoff-Baym equations by replacing $\Sigma$, but is not equal to the self energy for the Falicov-Kimball system in DMFT.  The Green's functions $G_1$ and $G_2$
correspond to the full or empty $f$ states after integrating out the
$f$ electrons. In the Falicov-Kimball model on the Bethe lattice, the
self consistency for the hybridization function $\Delta$ is simply a
linear combination of $G_1$ and $G_2$, so no nonlinear iteration is required in its numerical solution. This is a convenient simplification, but does not materially affect our algorithm.

In the first example, we consider a rapid ramp of the interaction
parameter $U(t)$, given by  
\beq{
  U(t)=\frac{U_0+U_1}{2}+\frac{U_1-U_0}{2}\erf\paren{5.922 (2t-1)}.
}
$U$ starts in the metallic phase $U_0=1$ at inverse temperature $\beta=5$,
and smoothly increases deep into the insulator transition $U_1=8$.
Experiments for various choices of $U_0$ and $U_1$ confirm that the
significant compressibility of the solution which we will demonstrate
for this case is typical.

\begin{figure}[t]
  \centering
  \begin{subfigure}[t]{.23\textwidth}
    \centering
    \includegraphics[width=\linewidth]{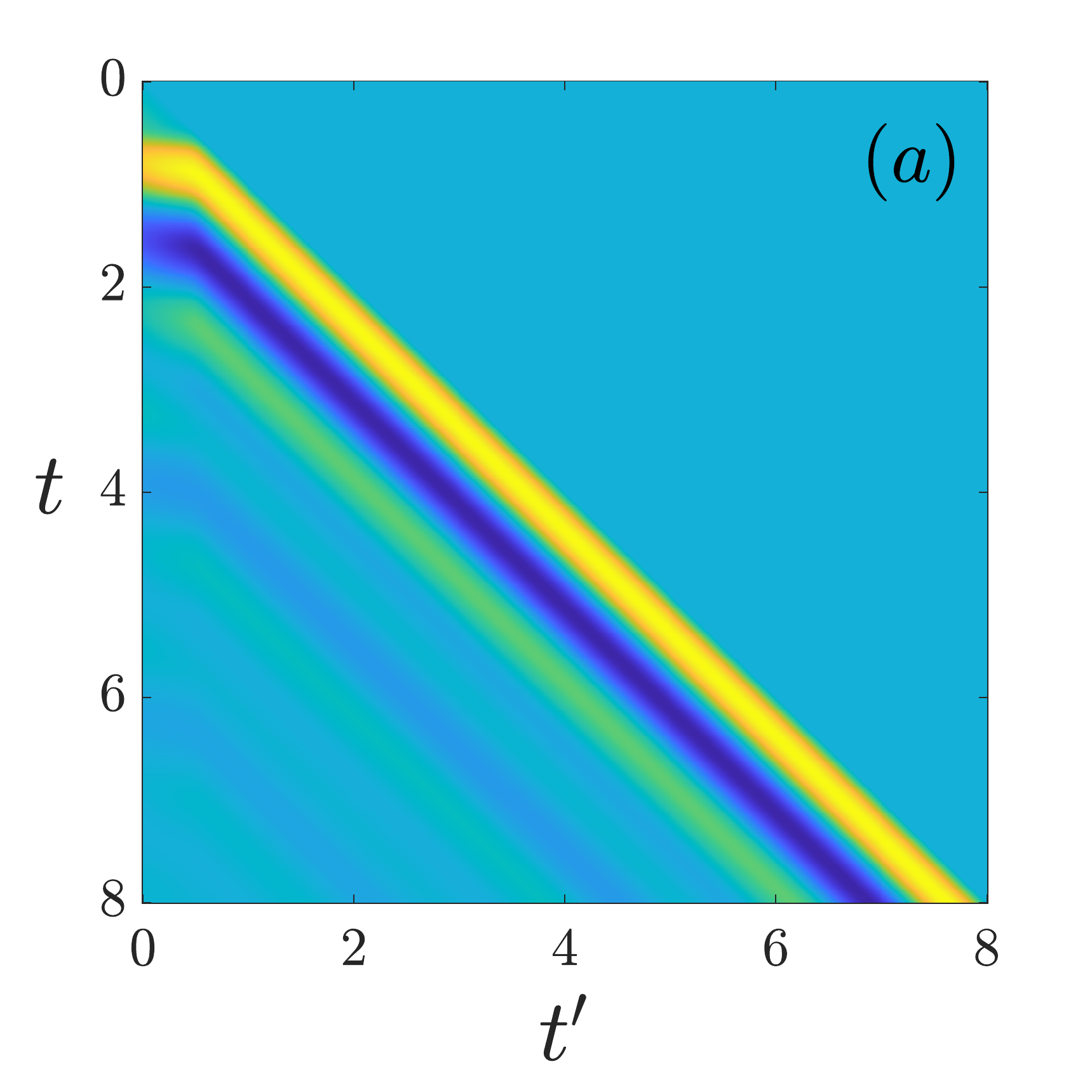}
    \captionlistentry{}
    \label{fig:gr1_ramp}
  \end{subfigure}
  \begin{subfigure}[t]{.23\textwidth}
    \centering
    \includegraphics[width=\linewidth]{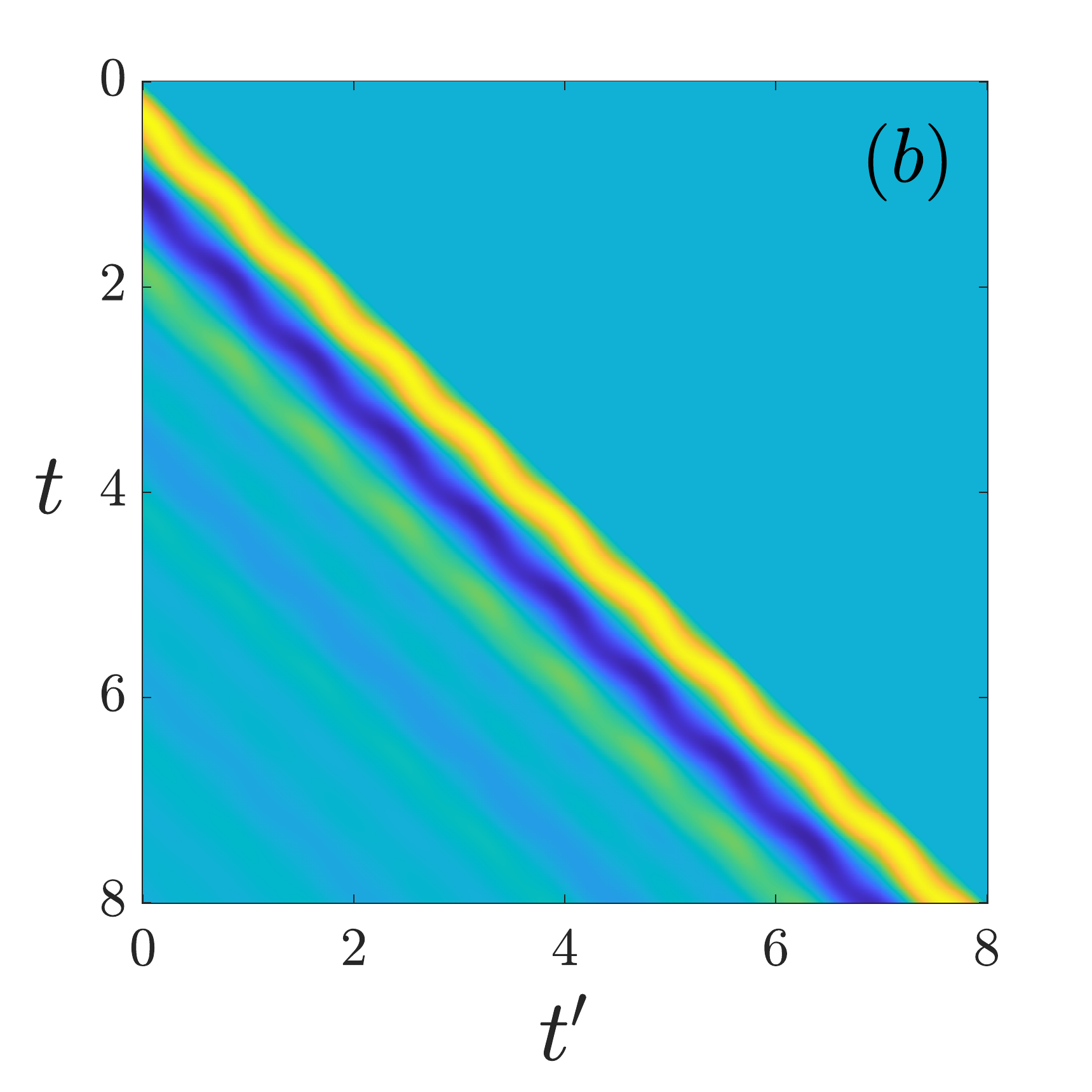}
    \captionlistentry{}
    \label{fig:gr1_floq}
  \end{subfigure}
  \vspace{-3ex}

  \caption{Plots of $G^R_1(t,t')$ for the (a) ramp and (b) Floquet examples of
  the Falicov-Kimball model, with propagation time $T = 8$. Note that the $t$ axis has been reversed to  reflect the matrix representation which we work with in the previous
  sections.}

\label{fig:gr1}
\end{figure}

In the second example, which we refer to as the Floquet example, we consider a periodic driving of
system parameters. Such protocols have been studied extensively in the
setting of Floquet
engineering~\cite{oka2019,bukov2015,mikami2016brillouin,rudner2020},
Floquet
prethermalization~\cite{peronaci2018,herrmann2018,abanin2015,machado2020long},
and high-harmonic
generation~\cite{sandholzer2019quantum,murakami2018,takayoshi2019}. In
particular, we simulate periodic driving of the Coulomb interaction,
\beq{
U(t)=U_{\text{eq}}+U_{\text{dr}} \sin(\omega t),
}
where $U_{\text{eq}}$ is the equilibrium interaction, $U_{\text{dr}}$ is
the driving strength, and $\omega$ is the driving frequency. We start in
the insulating phase $U_{\text{eq}}=8$ at the inverse temperature
$\beta=5$ and choose a resonant excitation $\omega=U_{\text{eq}}$ with
strength $U_{\text{dr}}=2$. As in the ramp example, our experiments show that
other parameter choices yield similar compressibility results. Plots of
$G_1^R(t,t')$ for the two examples with propagation time $T
= 8$ are shown in Fig. \ref{fig:gr1}.

We march the integro-differential equations in time using the implicit
trapezoidal rule, with history sums also discretized using the trapezoidal rule
as in Section \ref{sec:directmethod}. It is not our intention
here to obtain high accuracy calculations, and we use a low-order
discretization for simplicity, but an extension to high-order time
stepping methods and quadrature rules for history summation is
straightforward \cite{schuler2020}. $G^M$ is computed in advance to near
machine precision using a Legendre polynomial--based spectral method
with fixed point iteration, and then evaluated on the
equispaced $\tau$ grid \cite{gull2018chebyshev,dong2020}. Codes were written in Fortan using OpenBLAS for matrix
operations and linear algebra, including the evaluation of history sums
in the direct method, and FFTW was used for the FFTs in the fast
evaluation of the sums $I_{m,k}^{\rceil,2}$ \cite{xianyi12,openblas,frigo05}. All numerical experiments were performed on a laptop with an
Intel Xeon E-2176M 2.7GHz processor.

In the following experiments, errors are always computed as the maximum
entrywise difference between a computed solution and a reference. We
note that this is not the norm in which the TSVD guarantees
accuracy $\varepsilon$, but that is a minor technical issue and does not prevent
effective error control. When we present $\varepsilon$-ranks associated
with $G^R$ or $G^<$, we always maximize the rank over all blocks in the
HODLR partition, and simply refer to this as the rank of the compressed
representation. In all cases presented, for $G^R$ and $G^<$, this
coincides with the $\varepsilon$-rank of the largest block in the
partition. The number $L$ of levels can be adjusted to balance
computational cost and memory usage, but for all experiments, we simply
choose it so that the smallest blocks in the HODLR
partition are approximately $16 \times 16$.

\begin{figure}[t]
  \centering
  \begin{subfigure}[t]{.23\textwidth}
    \centering
    \includegraphics[width=\linewidth]{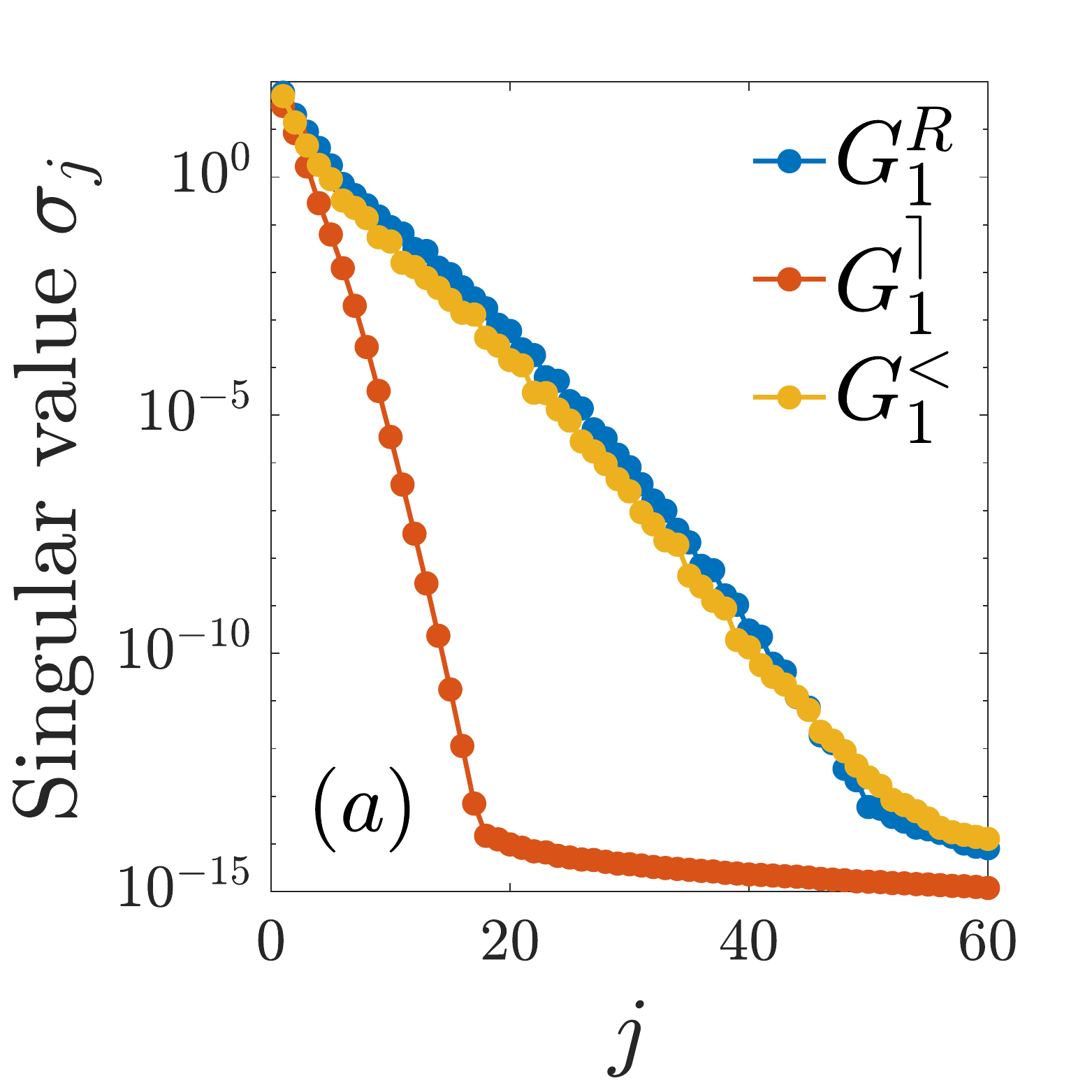}
    \captionlistentry{}
    \label{fig:varyeps_sval_ramp}
  \end{subfigure}
  \begin{subfigure}[t]{.23\textwidth}
    \centering
    \includegraphics[width=\linewidth]{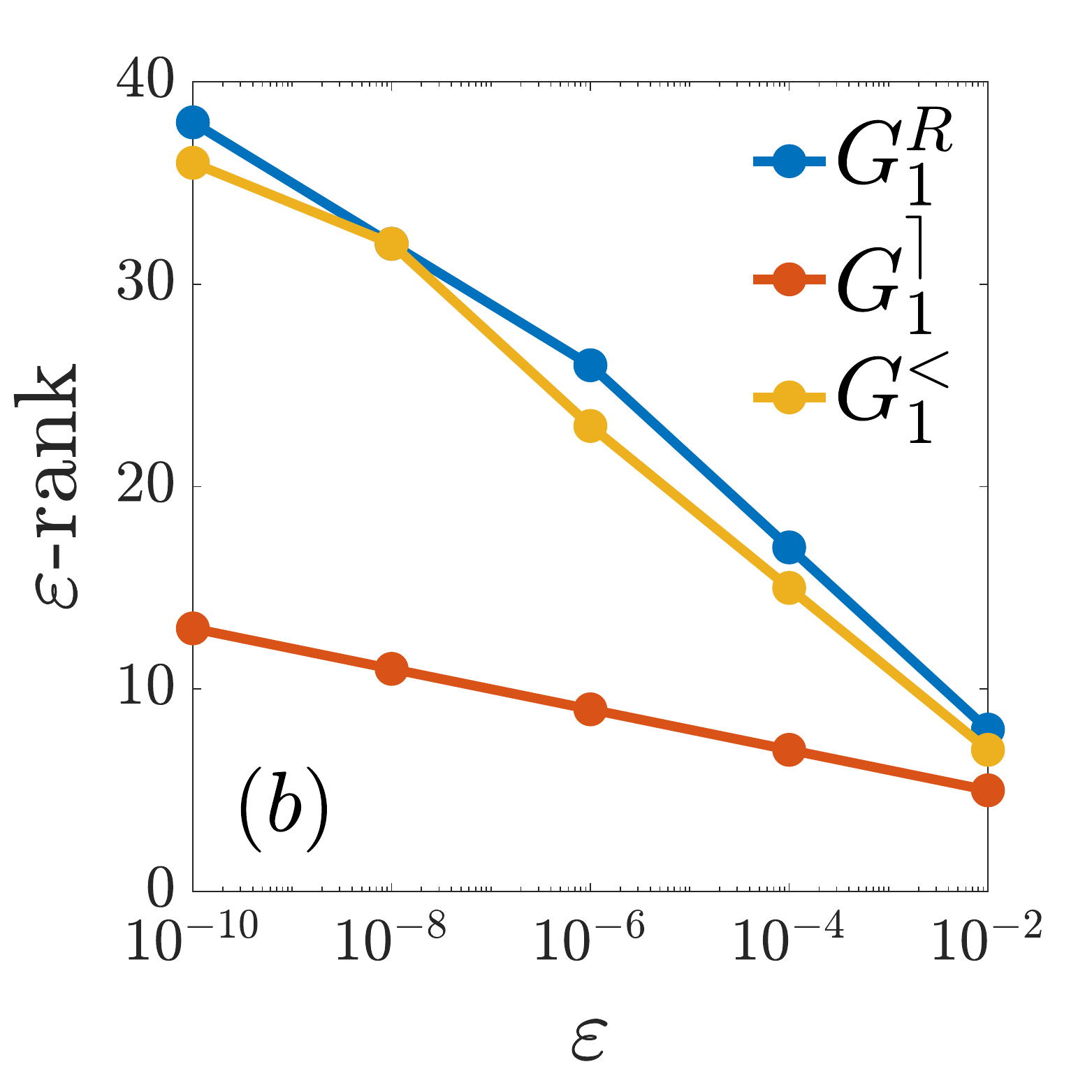}
    \captionlistentry{}
    \label{fig:varyeps_rank_ramp}
  \end{subfigure}
  \vspace{-3ex}

  \begin{subfigure}[t]{.23\textwidth}
    \centering
    \includegraphics[width=\linewidth]{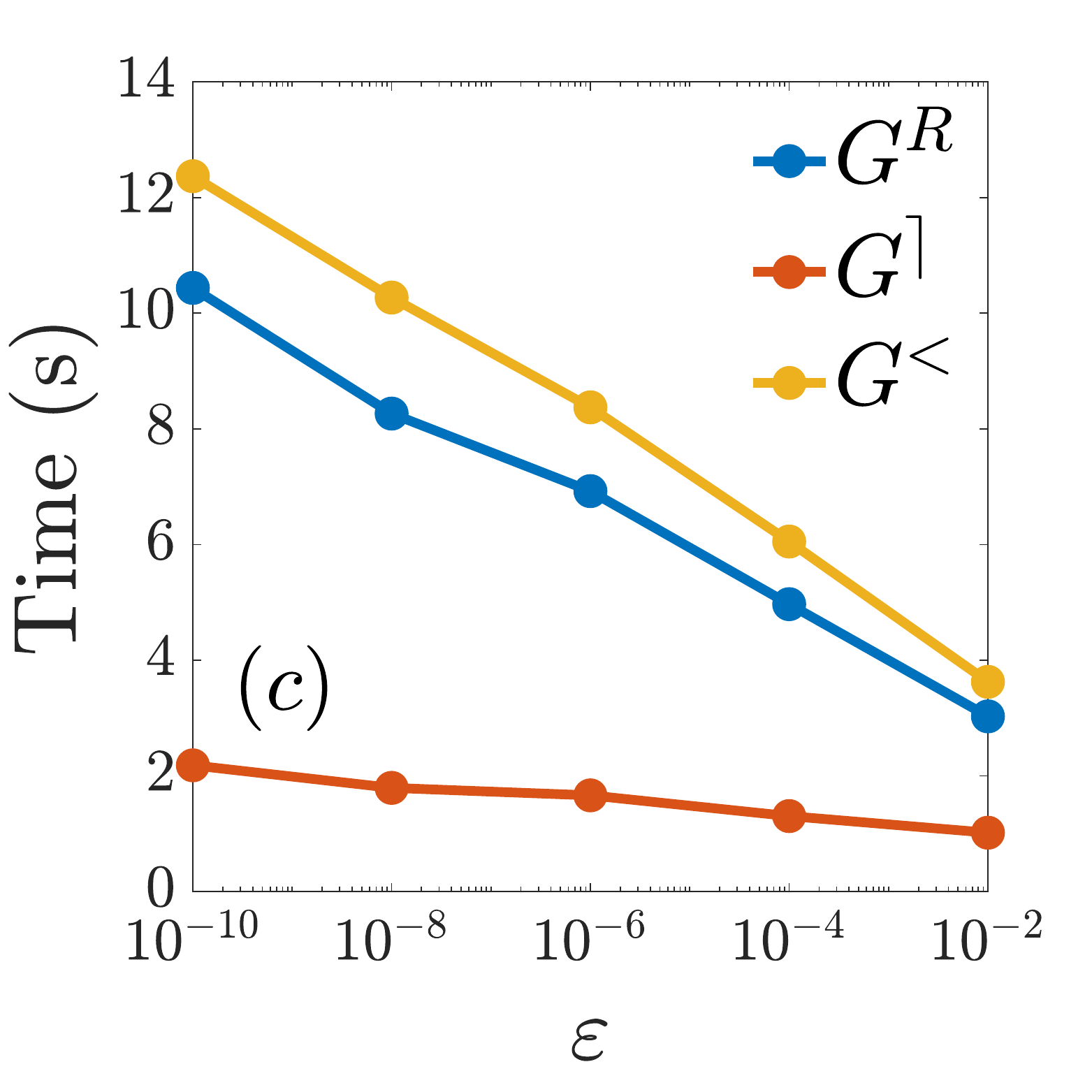}
    \captionlistentry{}
    \label{fig:varyeps_time_ramp}
  \end{subfigure}
  \begin{subfigure}[t]{.23\textwidth}
    \centering
    \includegraphics[width=\linewidth]{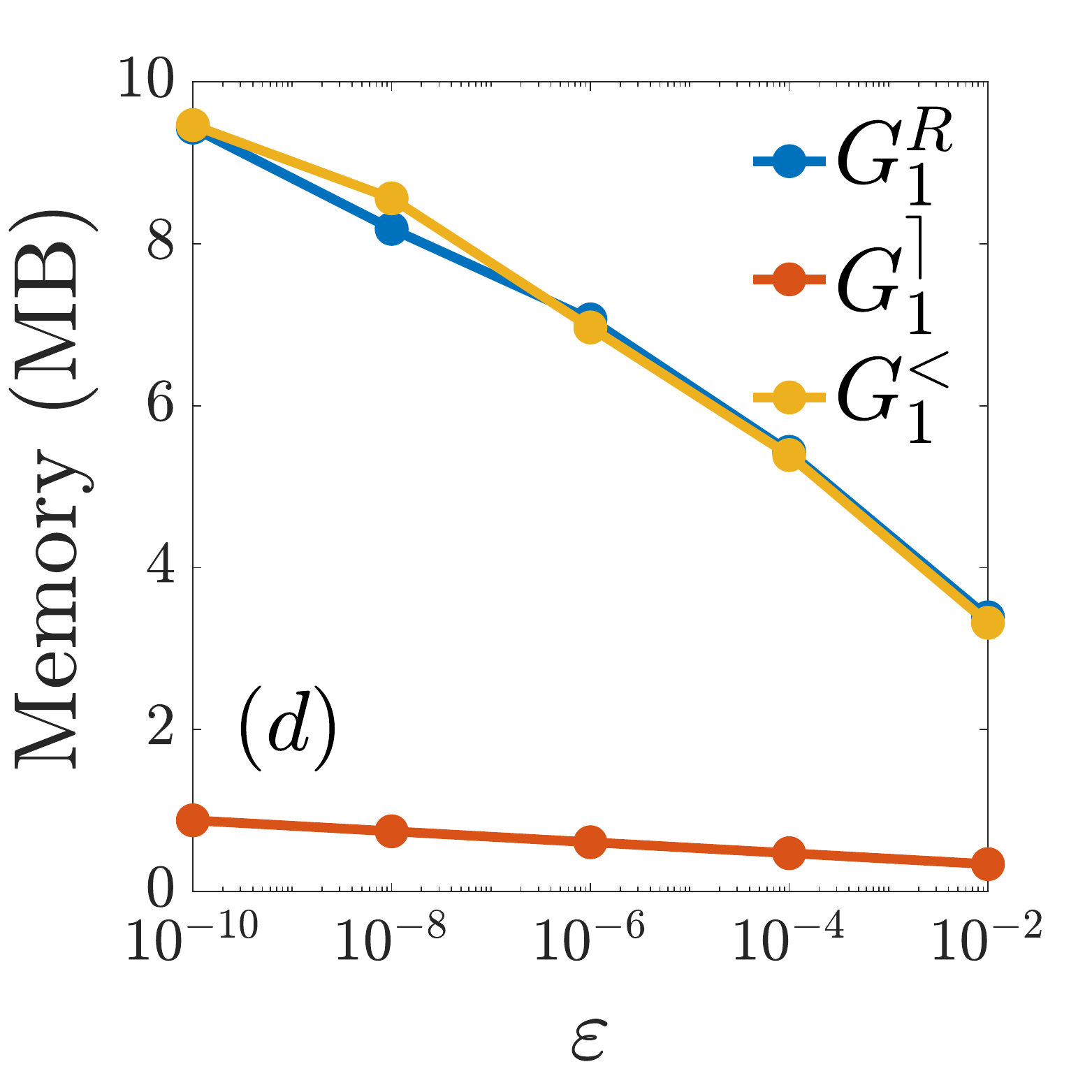}
    \captionlistentry{}
    \label{fig:varyeps_mem_ramp}
  \end{subfigure}
  \vspace{-3ex}

  \caption{Varying the SVD truncation tolerance $\varepsilon$ for the
  ramp example with propagation time $T = 64$, corresponding to $N =
  4096$ time steps. We plot (a) singular values of the largest
  block in the hierarchical partition for $G_1^R$ and $G_1^<$, and of
  $G_1^\rceil$, (b) the ranks of the compressed representations for $G_1$, (c) the time to
    compute each component of $G$, and (d) the memory required to store each component of $G_1$ in
    compressed form.}
\label{fig:varyeps_ramp}
\end{figure}

\begin{figure}[t]
  \centering
  \begin{subfigure}[t]{.23\textwidth}
    \centering
    \includegraphics[width=\linewidth]{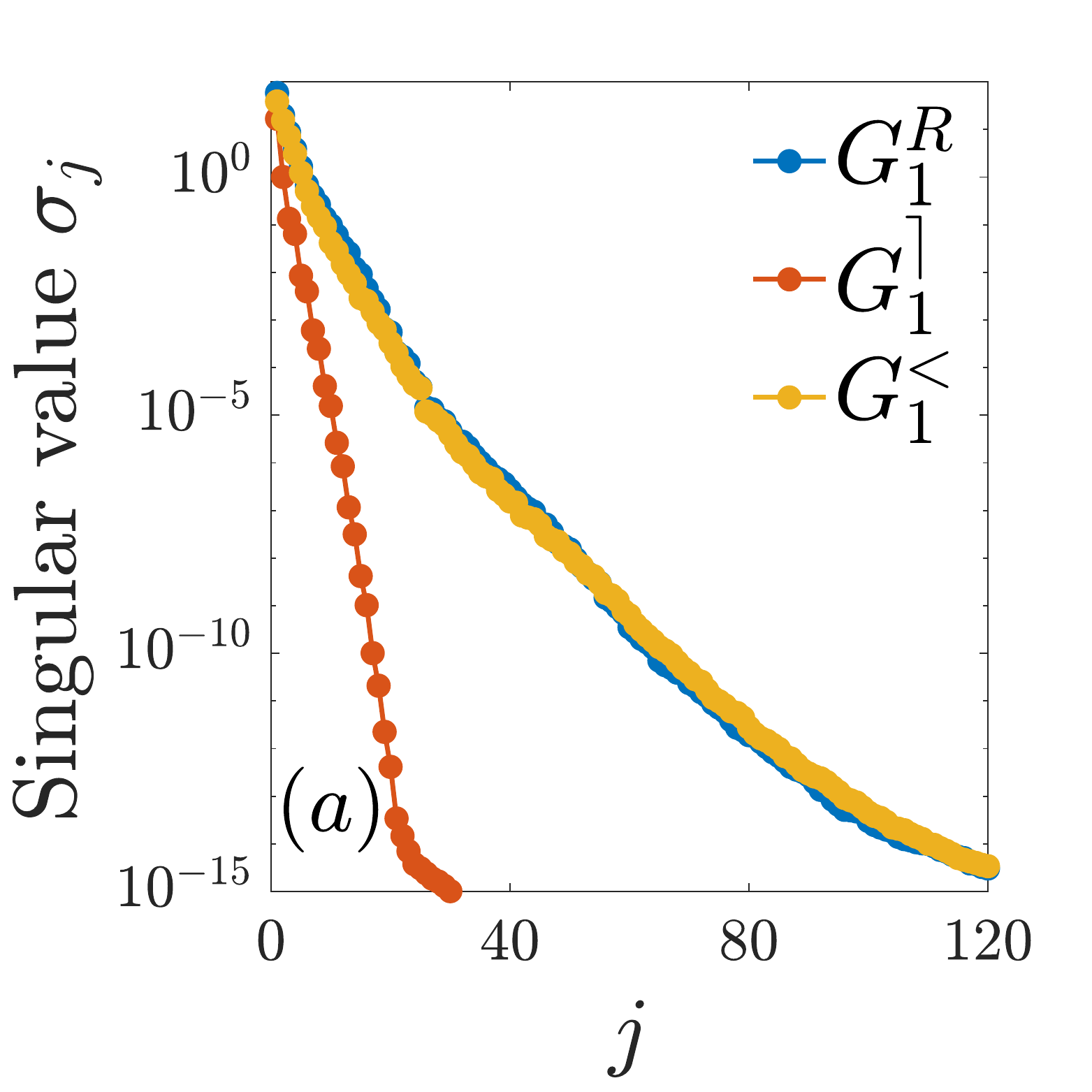}
    \label{fig:varyeps_sval_floq}
  \end{subfigure}
  \begin{subfigure}[t]{.23\textwidth}
    \centering
    \includegraphics[width=\linewidth]{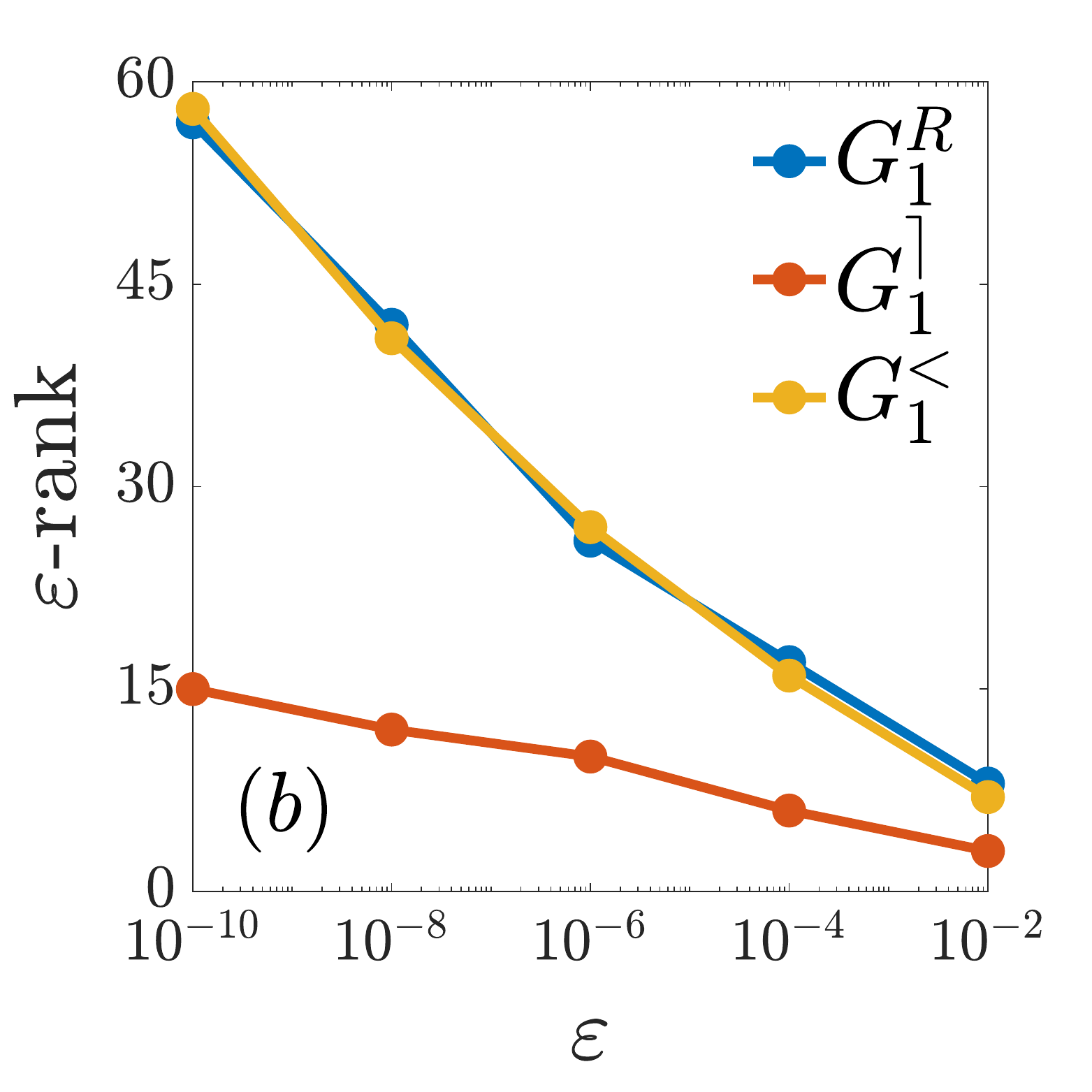}
    \label{fig:varyeps_rank_floq}
  \end{subfigure}
  \vspace{-3ex}

\begin{subfigure}[t]{.23\textwidth}
    \centering
    \includegraphics[width=\linewidth]{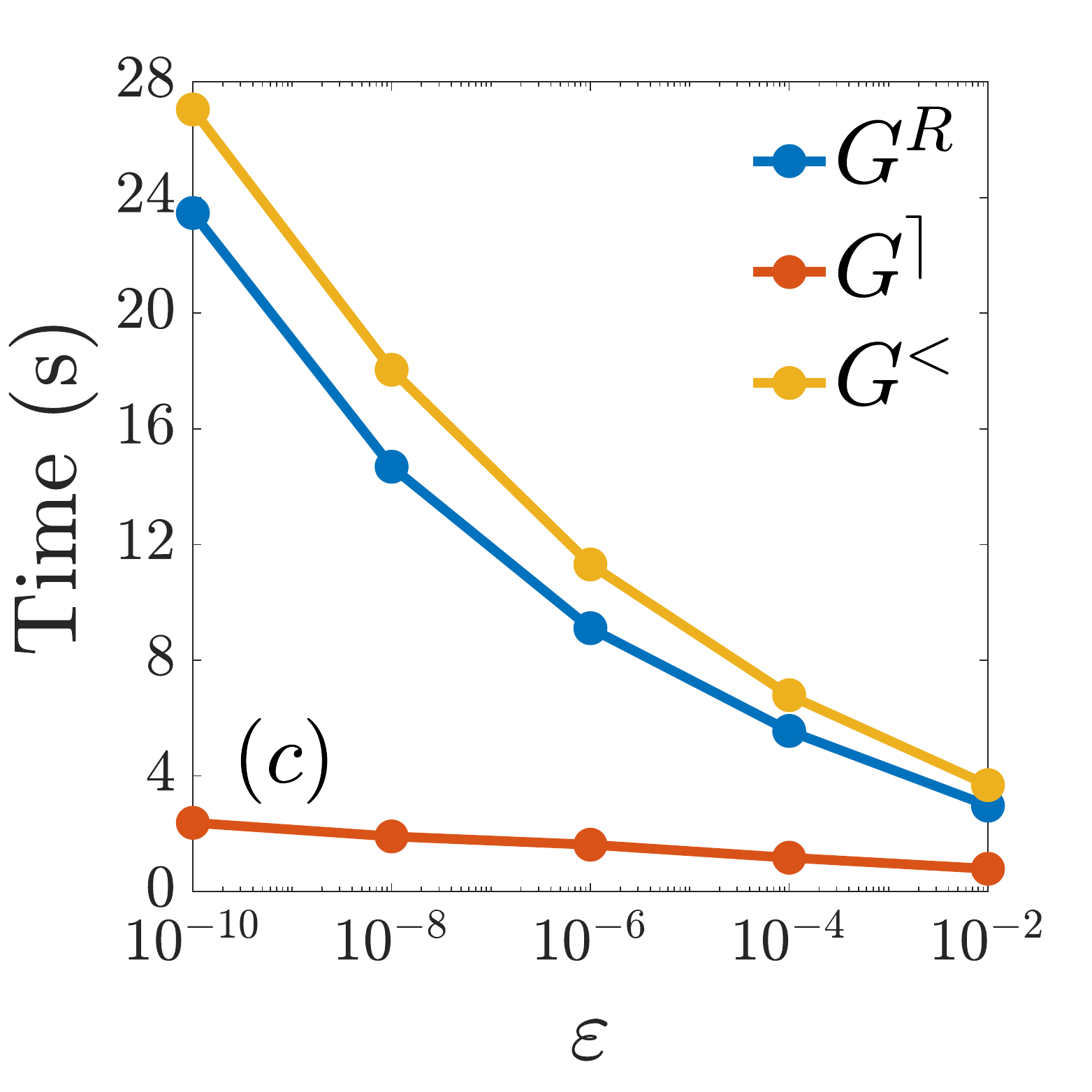}
    \label{fig:varyeps_time_floq}
  \end{subfigure}
  \begin{subfigure}[t]{.23\textwidth}
    \centering
    \includegraphics[width=\linewidth]{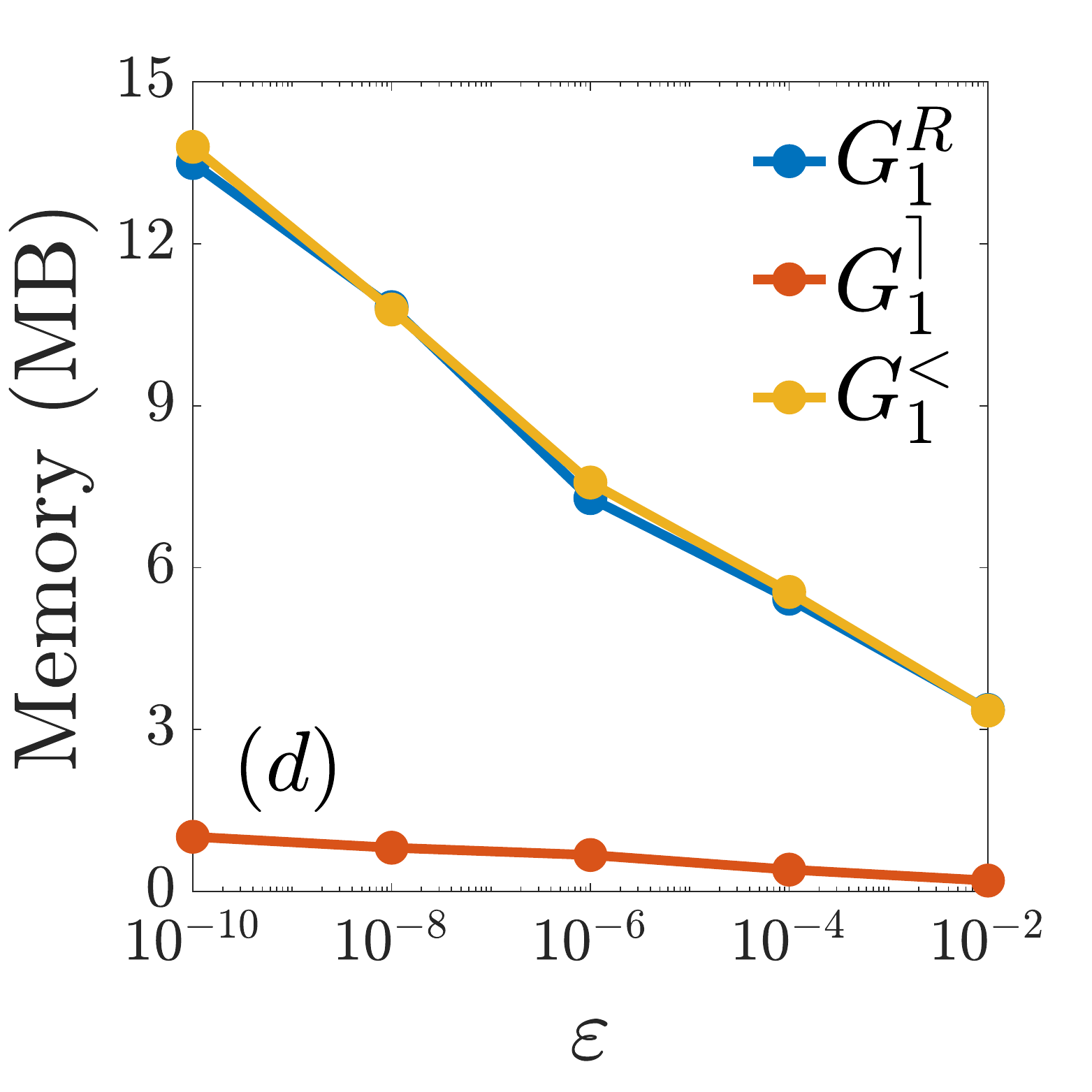}
    \label{fig:varyeps_mem_floq}
  \end{subfigure}
  \vspace{-3ex}

  \caption{Varying the SVD truncation tolerance $\varepsilon$ for the
  Floquet example with propagation time $T = 64$, corresponding to $N =
  4096$ time steps.The plots are analogous to those shown for the   ramp example in Fig. \ref{fig:varyeps_ramp}.}

\label{fig:varyeps_floq}
\end{figure}

We first examine the behavior of our method as the SVD truncation
tolerance $\varepsilon$ is varied, using both examples with the
propagation time $T = 64$, corresponding to $N = 4096$ time steps, and $M = 128$
Matsubara time steps. Errors compared with the direct method,  maximized
over all Keldysh components of the Green's function, are given for several values of $\varepsilon$ in Table
\ref{tab:varyeps}. For each experiment, the error is less than
$\varepsilon$. In Fig. \ref{fig:varyeps_sval_ramp}, we plot the singular
values of the largest blocks in the HODLR partitions of $G_1^R$ and
$G_1^<$, and of $G_1^\rceil$, for the ramp example. The singular values decay
approximately exponentially, and as a result, the $\varepsilon$-ranks of
these blocks increase only as approximately $\log \paren{1/\varepsilon}$, as
shown in Fig. \ref{fig:varyeps_rank_ramp}. The wall clock time required
to compute each component of $G$, shown in Fig.
\ref{fig:varyeps_time_ramp}, increases slightly more slowly than the
expected asymptotic $k \sim \log \paren{1/\varepsilon}$ rate for
these parameters. The memory required to store each component of $G_1$
is shown in Fig. \ref{fig:varyeps_mem_ramp}, and reflects the variation
in ranks seen in Fig. \ref{fig:varyeps_rank_ramp}. The results for $G_2$ are similar. Fig. \ref{fig:varyeps_floq}
contains analogous results for the Floquet example. Here, the decay of
the singular values, while still rapid, is slightly slower than
exponential, and this is reflected in the ranks, timings, and memory
usage.

\begin{table}[t]
  \centering
  \begin{tabular}{|c|c|c|c|c|c|c|}
    \hline
    \multicolumn{2}{|c|}{$\varepsilon$} & $10^{-2}$ & $10^{-4}$ & $10^{-6}$ & $10^{-8}$ &
    $10^{-10}$ \\ \hline
    \multirow{2}{*}{Error} & Ramp & $6 \times 10^{-3}$ & $6 \times 10^{-5}$ & $6
    \times 10^{-7}$ & $5 \times 10^{-9}$ & $5 \times 10^{-11}$ \\
    \cline{2-7}
    & Floquet & $8 \times 10^{-3}$ & $6 \times 10^{-5}$ & $8
    \times 10^{-7}$ & $7 \times 10^{-9}$ & $6 \times 10^{-11}$ \\
    \hline
  \end{tabular}
  
  \caption{Error of the Green's functions compared with the direct
  method, maximized over all Keldysh components, for different $\varepsilon$.
  For both examples, $T = 64$, $N = 4096$, $M = 128$.}
  \label{tab:varyeps}
\end{table}

We next fix $T = 8$ and $\varepsilon = 10^{-4}$, and measure errors and
ranks for $\Delta t$ corresponding to $N = 64, 128,\ldots,8192$
time steps. The errors are measured against a well-resolved solution,
with $M$ increased until convergence. The results are given in Table
\ref{tab:varydt}. We observe the expected second-order convergence with
$\Delta t$, until the SVD truncation error is
reached. We also find that the ranks are nearly constant as $N$ is
increased. Indeed, once the solution is resolved by the grid, the block ranks
cannot increase significantly as $\Delta t$ is further refined. In the regime of fixed $T$ and increasing $N$, therefore, we
are guaranteed $\OO{N^2 \log N}$ scaling of the computational cost and
$\OO{N \log N}$ scaling of the memory usage.

\begin{table}[t]
  \begin{subtable}[h]{0.45\textwidth}
  \centering
  \begin{tabular}{|c|c|c|c|c|c|c|}
    \hline

    \multirow{2}{*}{$\Delta t$} & \multicolumn{3}{c|}{Errors} &
    \multicolumn{3}{c|}{Ranks} \\ \cline{2-7}

     & \rule{0pt}{2.1ex} $G^R$ & $G^\rceil$ & $G^<$ &
    $G^R$ & $G^\rceil$ & $G^<$ \\ \hline
    
    $1/8$ & $1.01 \times 10^{-1}$ & $5.06 \times 10^{-2}$ & $8.83 \times
    10^{-2}$ & 9 & 7 & 8 \\ \hline

    $1/16$ & $2.50 \times 10^{-2}$ & $1.32 \times 10^{-2}$ & $2.25 \times
    10^{-2}$ & 9 & 7 & 9 \\ \hline

    $1/32$ & $6.25 \times 10^{-3}$ & $3.32 \times 10^{-3}$ & $5.65 \times
    10^{-3}$ & 9 & 7 & 9 \\ \hline

    $1/64$ & $1.56 \times 10^{-3}$ & $8.29 \times 10^{-4}$ & $1.42 \times
    10^{-3}$ & 9 & 7 & 9 \\ \hline

    $1/128$ & $3.90 \times 10^{-4}$ & $2.05 \times 10^{-4}$ & $3.51 \times
    10^{-4}$ & 10 & 7 & 9 \\ \hline

    $1/256$ & $9.78 \times 10^{-5}$ & $4.95 \times 10^{-5}$ & $8.56 \times
    10^{-5}$ & 10 & 7 & 9 \\ \hline

    $1/512$ & $3.31 \times 10^{-5}$ & $4.79 \times 10^{-5}$ & $6.03 \times
    10^{-5}$ & 10 & 7 & 9 \\ \hline

    $1/1024$ & $3.61 \times 10^{-5}$ & $4.79 \times 10^{-5}$ & $5.09 \times
    10^{-5}$ & 11 & 7 & 10 \\ \hline

  \end{tabular}
  \caption{}
  \end{subtable}
  \begin{subtable}[h]{0.45\textwidth}
  \centering
  \begin{tabular}{|c|c|c|c|c|c|c|}
    \hline

    \multirow{2}{*}{$\Delta t$} & \multicolumn{3}{c|}{Errors} &
    \multicolumn{3}{c|}{Ranks} \\ \cline{2-7}

     & \rule{0pt}{2.1ex} $G^R$ & $G^\rceil$ & $G^<$ &
    $G^R$ & $G^\rceil$ & $G^<$ \\ \hline
    
    $1/8$ & $1.09 \times 10^{-1}$ & $1.07 \times 10^{-1}$ & $1.07 \times
    10^{-1}$ & 9 & 6 & 8 \\ \hline

    $1/16$ & $2.81 \times 10^{-2}$ & $2.79 \times 10^{-2}$ & $2.79 \times
    10^{-2}$ & 9 & 6 & 7 \\ \hline

    $1/32$ & $7.08 \times 10^{-3}$ & $7.02 \times 10^{-3}$ & $7.03 \times
    10^{-3}$ & 9 & 6 & 8 \\ \hline

    $1/64$ & $1.78 \times 10^{-3}$ & $1.76 \times 10^{-3}$ & $1.77 \times
    10^{-3}$ & 10 & 6 & 9 \\ \hline

    $1/128$ & $4.50 \times 10^{-4}$ & $4.40 \times 10^{-4}$ & $4.45 \times
    10^{-4}$ & 10 & 6 & 9 \\ \hline

    $1/256$ & $1.22 \times 10^{-4}$ & $1.11 \times 10^{-4}$ & $1.20 \times
    10^{-4}$ & 10 & 6 & 9 \\ \hline

    $1/512$ & $3.74 \times 10^{-5}$ & $5.65 \times 10^{-5}$ & $5.65 \times
    10^{-5}$ & 10 & 6 & 9 \\ \hline

    $1/1024$ & $3.56 \times 10^{-5}$ & $5.94 \times 10^{-5}$ & $5.94 \times
    10^{-5}$ & 11 & 6 & 9 \\ \hline

  \end{tabular}
  \caption{}
  \end{subtable}

  \caption{Errors and ranks with a varying time step $\Delta t$ and fixed
  propagation time $T = 8$,
  for the (a) ramp and (b) Floquet examples. Errors and ranks for each component are maximized over
  $G_1$ and $G_2$.}
  \label{tab:varydt}
\end{table}

The more challenging regime is that of increasing $T$ and $N$ with fixed
time step $\Delta t$. We take $\Delta t = 1/64$, and doubling values of
the propagation time $T = 4, 8, \ldots, 4096$, corresponding to 
$N = 256, 512, 1024, \ldots, 262\,144$ time steps.
We note that in our experiments, the maximum error for fixed $\Delta t$
is observed to be approximately constant as $N$ and $T$ are increased,
so according to Table \ref{tab:varydt} these simulations have
approximately three-digit accuracy. We fix $M =
128$, which is sufficient to eliminate it as a dominant source of error,
and $\varepsilon = 10^{-4}$.

\begin{figure}[t]
  \centering
  \begin{subfigure}[t]{.23\textwidth}
    \centering
    \includegraphics[width=\linewidth]{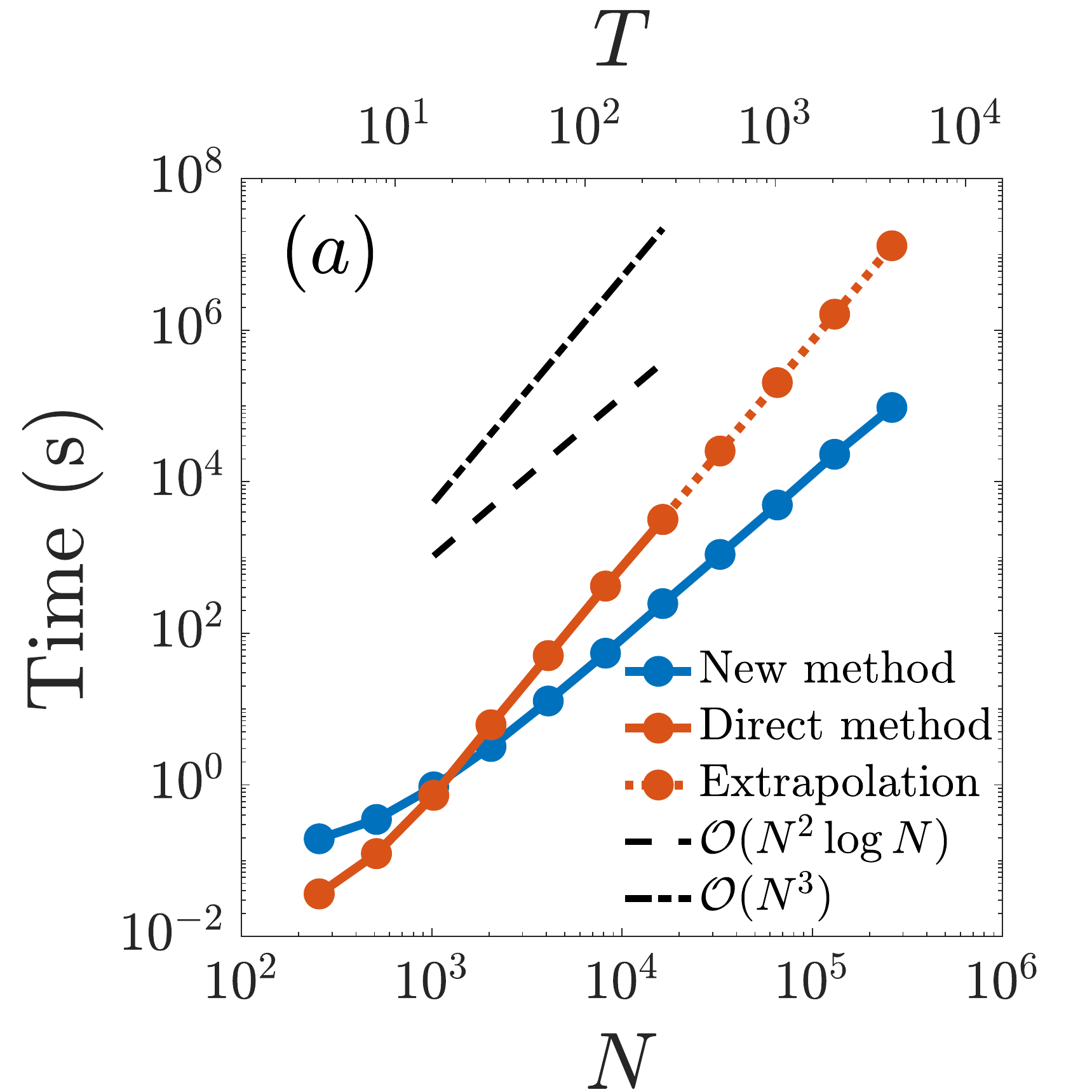}
    \captionlistentry{}
    \label{fig:varyt_time_floq}
  \end{subfigure}
  \begin{subfigure}[t]{.23\textwidth}
    \centering
    \includegraphics[width=\linewidth]{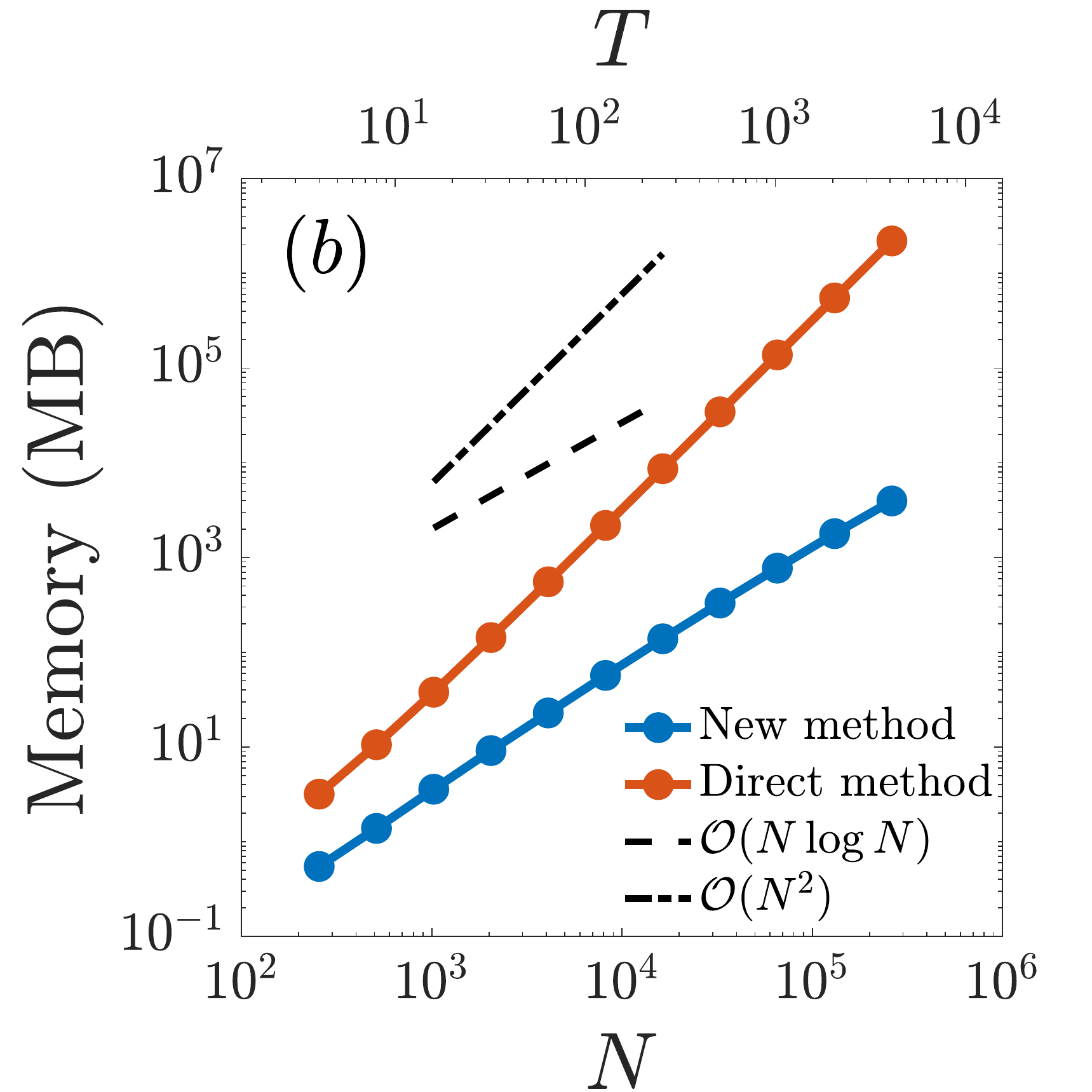}
    \captionlistentry{}
    \label{fig:varyt_mem_floq}
  \end{subfigure}
  \vspace{-3ex}

  \caption{Increasing the propagation time $T$ with $N$ fixed time steps
  of size $\Delta t$ and SVD truncation tolerance $\varepsilon =
  10^{-4}$, for the Floquet example. The plots are analogous to those shown for the ramp example in Fig. \ref{fig:varyt_ramp}.}

\label{fig:varyt_floq}
\end{figure}

Fig. \ref{fig:varyt_ramp} shows the
time and memory required for each
simulation for the ramp example, using our algorithm and the direct
method. For sufficiently large values of $N$, the direct method becomes
impractical, and we obtain timings by extrapolation. We observe the expected scalings. For the largest simulation,
which has $T = 4096$ and $N = 262\,144$, our algorithm takes approximately
$26.5$ hours and uses $3.8$ GB of memory to store the Green's functions, whereas the direct
method would take approximately $5$ months and use $2.2$ TB of memory.
This implies a speedup factor of
approximately $135$, and a compression factor of approximately $580$. A simulation which would take the direct method
$24$ hours, at $N \approx 49250$, using $78$ GB of memory,
would take our method approximately
$42$ minutes and use $512$ MB of memory. Our method is faster whenever $N
> 1200$, and it uses less memory for all values of $N$. The results for
the Floquet example, given in Fig. \ref{fig:varyt_floq}, are nearly
identical.

\begin{figure}[t]
  \centering
  \begin{subfigure}[t]{.23\textwidth}
    \centering
    \includegraphics[width=\linewidth]{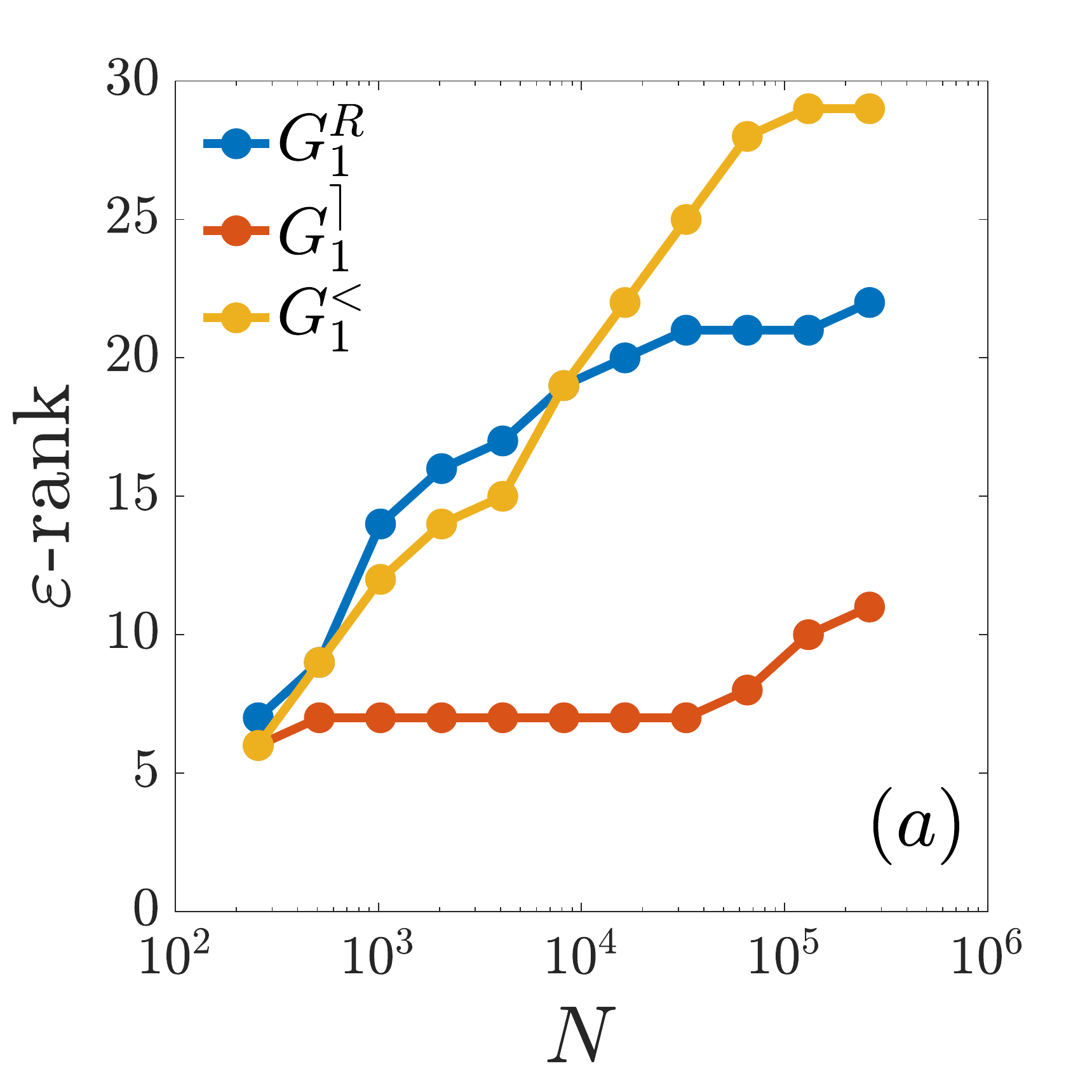}
    \captionlistentry{}
    \label{fig:varyt_rank_ramp}
  \end{subfigure}
  \vspace{-3ex}
  \begin{subfigure}[t]{.23\textwidth}
    \centering
    \includegraphics[width=\linewidth]{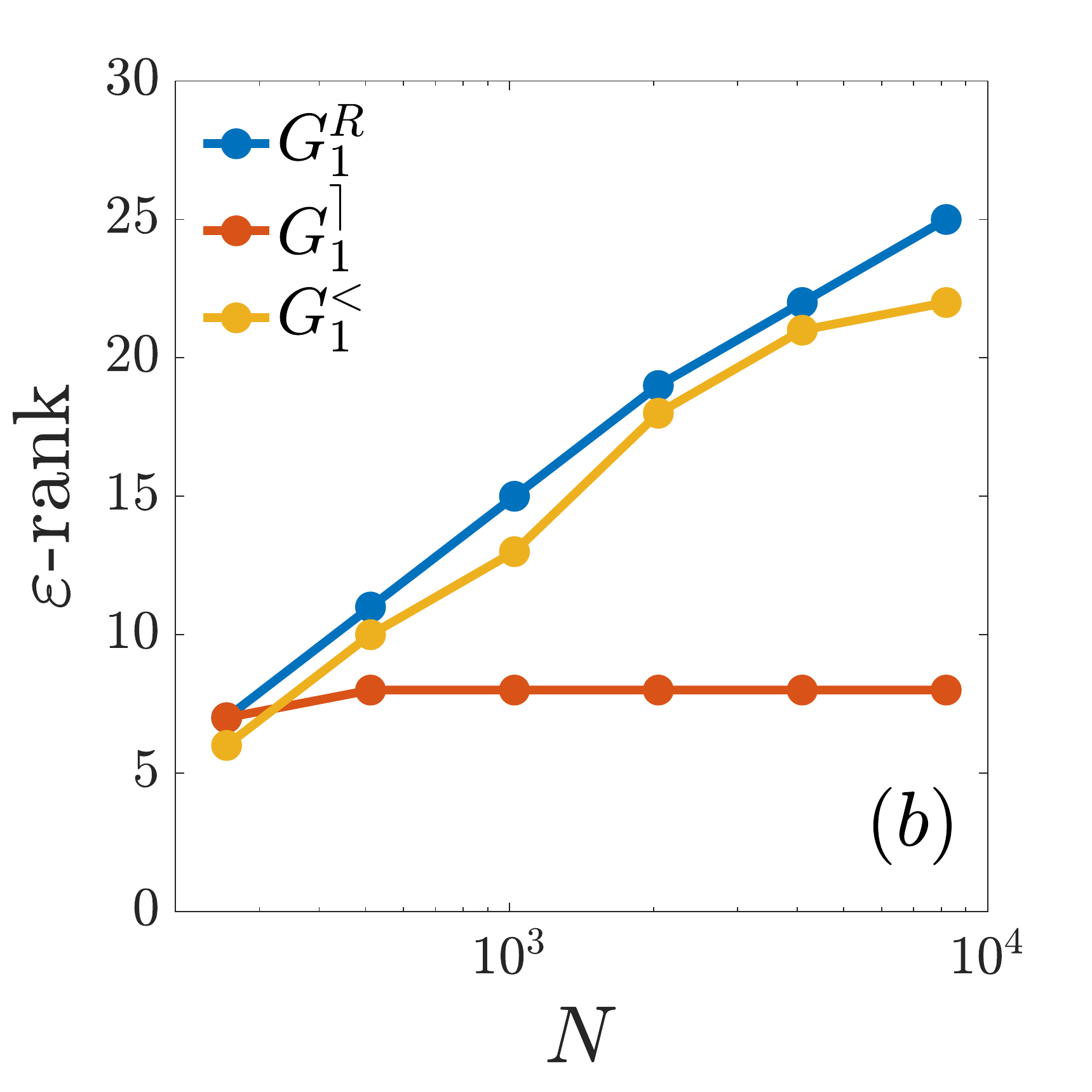}
    \captionlistentry{}
    \label{fig:offlineranksfk_ramp}
  \end{subfigure}
  \vspace{-3ex}
  \centering
  \begin{subfigure}[t]{.23\textwidth}
    \centering
    \includegraphics[width=\linewidth]{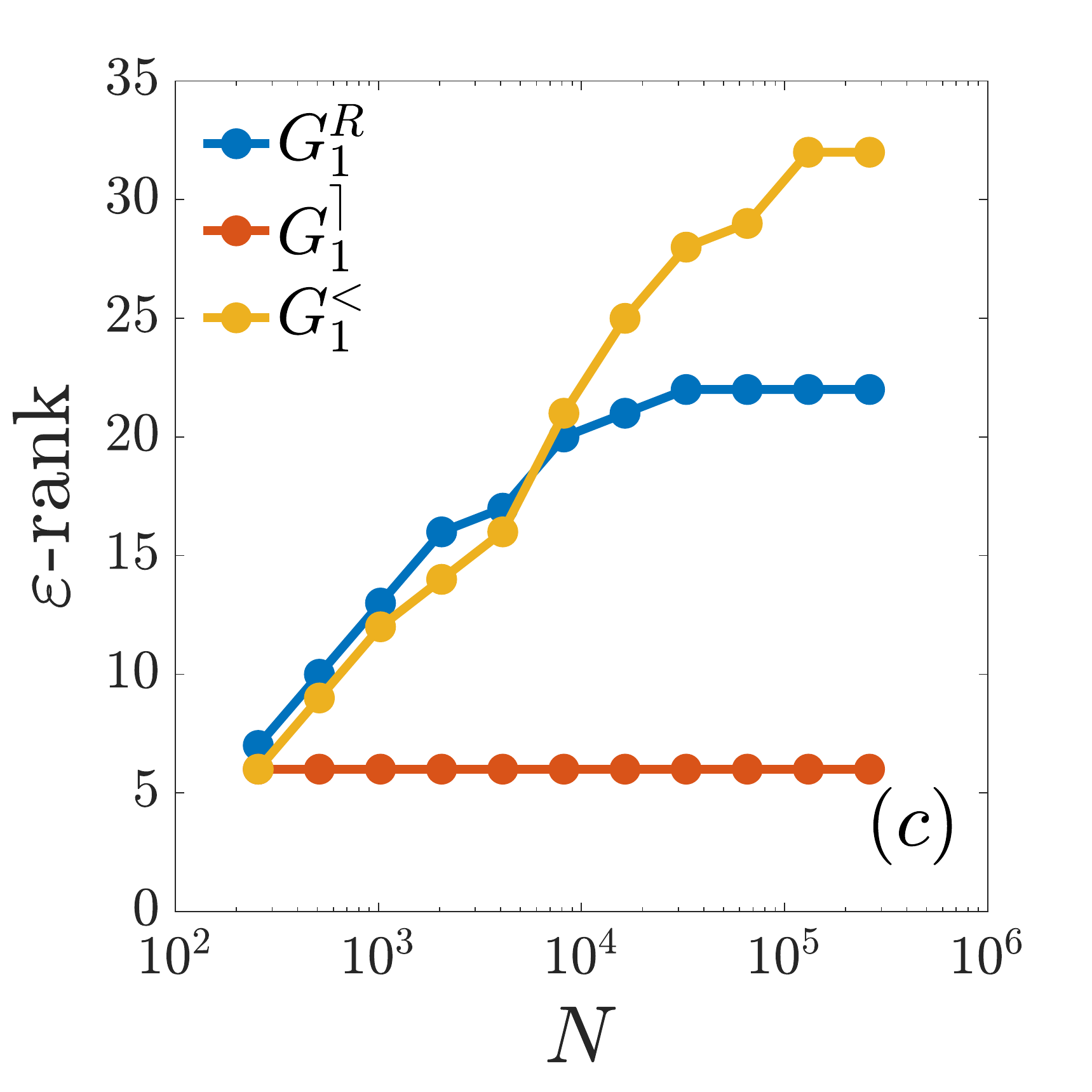}
    \captionlistentry{}
    \label{fig:varyt_rank_floq}
  \end{subfigure}
  \begin{subfigure}[t]{.23\textwidth}
    \centering
    \includegraphics[width=\linewidth]{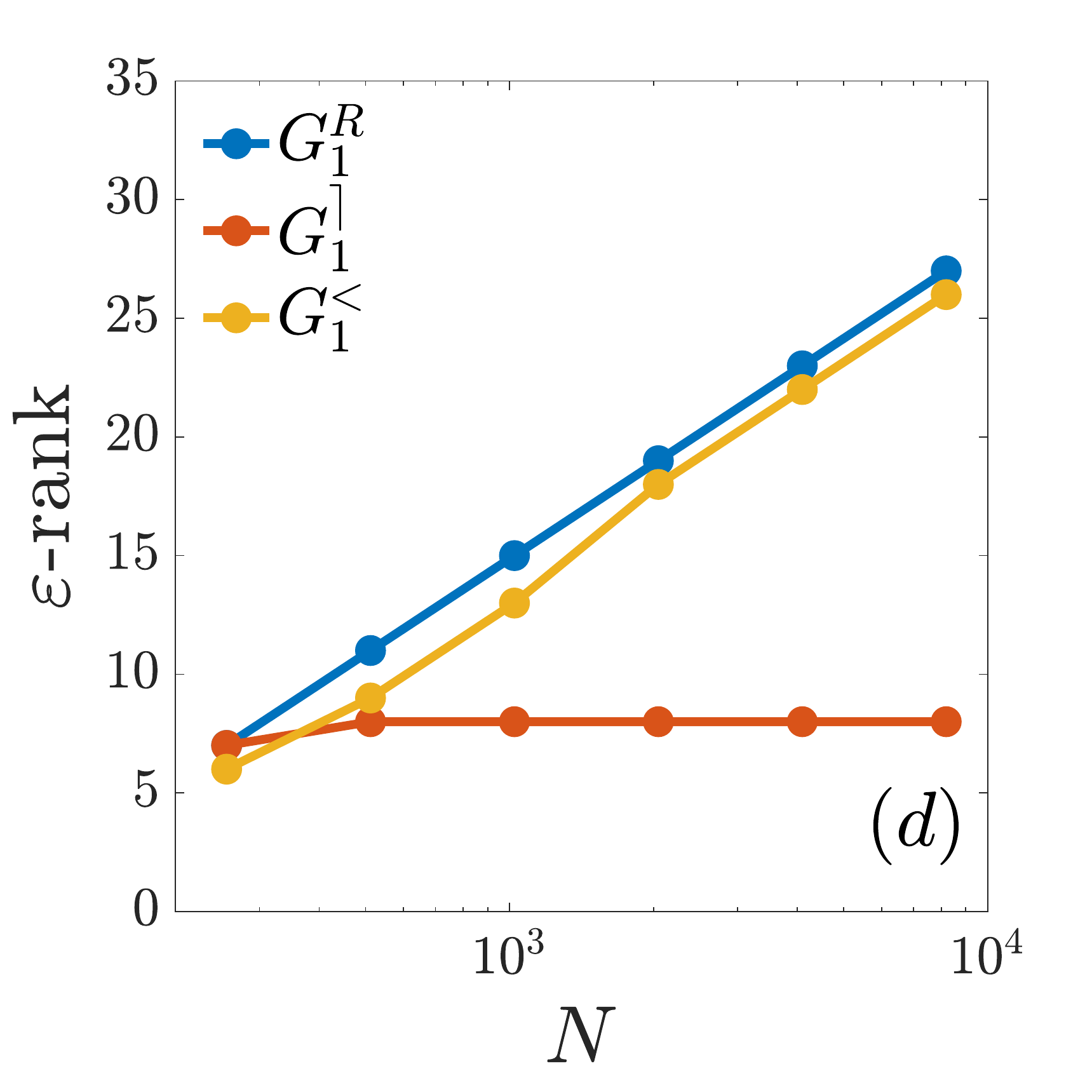}
    \captionlistentry{}
    \label{fig:offlineranksfk_floq}
  \end{subfigure}
  \vspace{-3ex}

  \caption{Ranks of the compressed representations for $G_1$ with
  increasing propagation times $T$ corresponding to $N$ time steps of
  fixed size $\Delta t$, and 
  SVD truncation tolerance $\varepsilon =
  10^{-4}$, for the ramp (first row, (a) \& (b)), and Floquet (second
  row, (c) \& (d)) examples. The first column ((a) \& (c)) contains
  ranks computed on the fly, and the second ((b) \& (d)) contains ranks
  computed offline for the smaller values of $N$.
   }
  \label{fig:varyt_rank}
\end{figure}

Rank information for the two examples is given in Figs.
\ref{fig:varyt_rank_ramp} and \ref{fig:varyt_rank_floq}, respectively.
The
crucial empirical observation enabling our complexity gains is that the maximum ranks grow at most logarithmically with $N$. 

\subsection{Hierarchical low rank compression in other systems}

We have demonstrated an implementation specialized for the
Falicov-Kimball problem for simplicity, but our method will be efficient
for any system in which the Green's functions and self energies are
HODLR compressible. This property can be tested offline,
for a solution which has already been computed, by simply measuring the
$\varepsilon$-ranks of the blocks in the compressed representations.
In this way, we can determine whether or not our algorithm will be
effective when applied to other systems of interest.

Let us first verify for the Falicov-Kimball model that an offline
measurement gives similar $\varepsilon$-ranks to those computed online using our SVD
update algorithm. This will show that the update algorithm is yielding
accurate results without requiring ranks which are larger than
necessary, and therefore that measuring $\varepsilon$-ranks offline is
sufficient to understand online performance. For both the ramp and
Floquet examples, we take the same parameters as in the previous
experiment, with $T = 4,8,\ldots,128$ and $N = 256,512,\ldots,8192$, and
compute $G_1$ by the direct method. Then, for $\varepsilon = 10^{-4}$,
we measure directly the same $\varepsilon$-ranks which were computed on the fly and
are shown in Figs. \ref{fig:varyt_rank_ramp} and
\ref{fig:varyt_rank_floq}; that
is, we simply compute the SVD of the blocks and count the number of
singular values above $\varepsilon$. The results are given in
Figs. \ref{fig:offlineranksfk_ramp} and \ref{fig:offlineranksfk_floq}.

Comparing the online and offline results shows that our
on the fly procedure in fact obtains smaller ranks than those computed
offline from the full solution. This observation merits some comment.
One would not expect the online and offline
$\varepsilon$-ranks to be identical, since the history sums used in the
online algorithm contain an error of magnitude $\varepsilon$. Our only concern would be if the
smaller online $\varepsilon$-ranks were accompanied by an error of magnitude much larger than the expected $\varepsilon$,
and this is not the case for any of the examples treated in Figs.
\ref{fig:offlineranksfk_ramp} and \ref{fig:offlineranksfk_floq}.
More generally, a difference in the $\varepsilon$-ranks of two matrices does not
imply a large difference between the matrices themselves, measured in
some norm; for example, the $n \times n$ diagonal matrices
$A=\text{diag}\paren{1,1.5\varepsilon,\cdots,1.5\varepsilon}$ and
$B=\text{diag}\paren{1,0.5\varepsilon,\cdots,0.5\varepsilon}$ have
$\varepsilon$-ranks $n$ and $1$, respectively, and
$\norm{A-B}_2/\norm{A}_2 = \varepsilon$. 
In our examples, compared with the size of the matrices which have been compressed, the
observed discrepancy in the ranks is in practice negligible.




We can now estimate the effectiveness of our method for other systems by
this offline procedure. As an example, we use the Hubbard
model, a paradigmatic problem in the theory of strongly correlated
systems which demonstrates a variety of phenomena, including the metal-insulator
transition and magnetic phases~\cite{mott1968metal,imada1998metal,georges1996dynamical,georges2004}.
The Hamiltonian,
\beq{
  H(t)=- J \sum_{\langle i,j\rangle,\sigma}  \exp^{-\I \phi(t)} c_{i\sigma}^{\dagger} c_{j\sigma} + U \sum_i (n_{i\uparrow}-1/2)(n_{i\downarrow}-1/2),
}
describes the competition between the kinetic energy and the
on-site Coulomb interaction.
Here, $c_{i\sigma}$ is the annihilation operator at site $i$ for
spin $\sigma$, $n_{i\sigma}$ is the corresponding density operator, $J$
is the hopping parameter, and $U$ is the Coulomb strength.
Coupling to an external electric field $E(t)$ is introduced via the Peierls
substitution and enters as a time-dependent phase of the hopping
parameter $\phi(t)=-l A(t)$, where $l$ is the lattice constant.
The vector potential $A$ is obtained from the electric field by
$A(t)=-\int_0^t d\bar t E(\bar t)$. We consider two
characteristic cases at half-filling: the weak coupling regime, in which
the bandwidth $W$ is larger the Coulomb interaction, $W>U$, and the
system is metallic, and the strong coupling regime, in which the Coulomb
interaction dominates, $U>W$, and the system is a Mott insulator.  

\paragraph{Correlated metal within GW}
In the weak coupling regime, we consider the GW approximation for the
self energy. This approximation has been used extensively for the
realistic modeling of molecules, weakly correlated extended systems,
coupling with bosonic excitations, and
screening~\cite{perfetto2018ultrafast,molina2017ab,sangalli2019many,marini2009yambo,deslippe2012,sentef2013,abdurazakov2018,kemper2018,sayyad2019}. In combination with DMFT, it was used in and out of equilibrium to study
plasmonic physics in correlated
systems~\cite{sun2002,ayral2013,boehnke2016strong,golevz2019multiband,golez2019,golez2017}.
We consider a one-dimensional setup with translational invariance and a
paramagnetic phase; see
Ref.~\onlinecite{schuler2020} for a detailed description. In this case,
the single particle energy includes the coupling to the external vector
potential as $\epsilon(k-A)=-2 J \cos(k-A)$.

We examine two excitation protocols. The first involves a short electric
field pulse, as typically used in pump-probe experiments, parametrized
by
\beq{
  E(t)=E_0 \sin\paren{\omega (t-t_0)} \exp\paren{-4.2(t-t_0)/t_0^2},
\label{eq:pulse}
}
where the delay $t_0=2\pi/\omega$ is chosen so that the pulse contains
one cycle. We use a pump strength $E_0 = 5$ and base frequency
$\omega = 4$. The second is a Floquet driving of the electric field,
\beq{
  E(t)=E_0^F \sin\paren{\omega_F t},
  \label{eq:floquet}
}
with driving strength $E_0^F = 1$ and frequency $\omega_F = 2$. In both
cases we fix the Coulomb strength $U=2$, and at equilibrium the inverse
temperature $\beta=20$.

\begin{figure}[t]
  \centering
  \begin{subfigure}[t]{.23\textwidth}
    \centering
    \includegraphics[width=\linewidth]{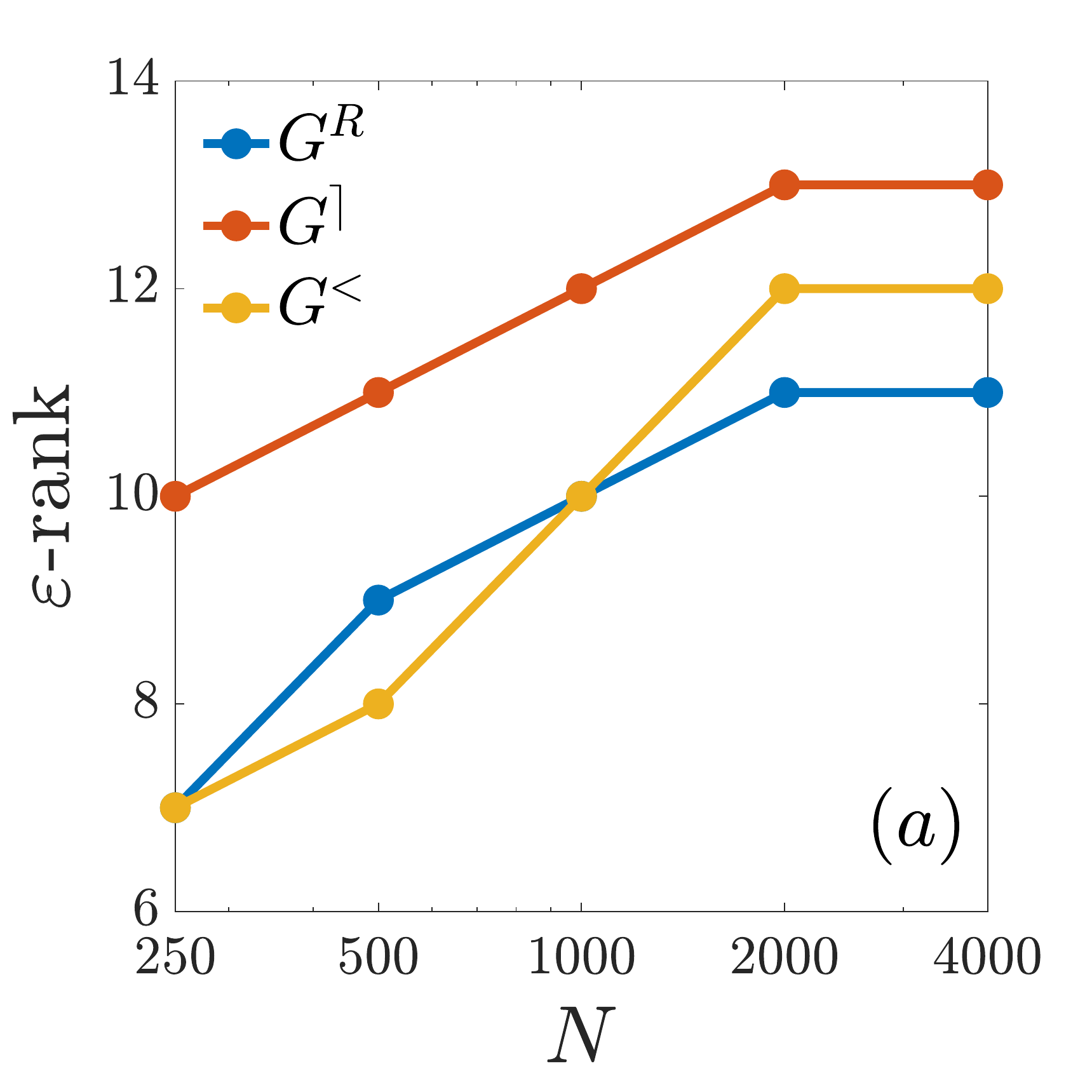}
    \captionlistentry{}
    \label{fig:nvrankoff_gwpulse}
  \end{subfigure}
  \begin{subfigure}[t]{.23\textwidth}
    \centering
    \includegraphics[width=\linewidth]{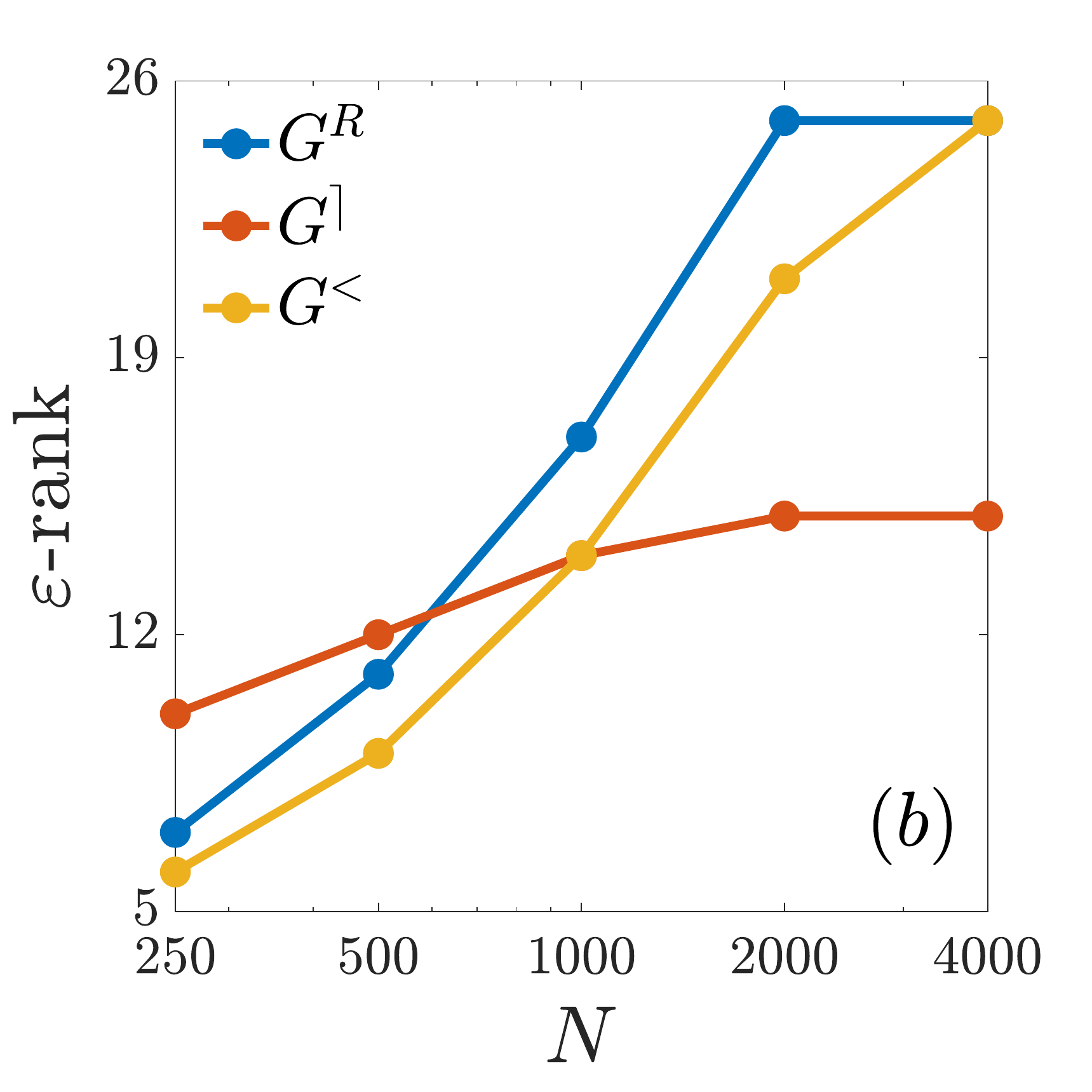}
    \captionlistentry{}
    \label{fig:nvrankoff_gwfloq}
  \end{subfigure}
  \vspace{-3ex}

\begin{subfigure}[t]{.23\textwidth}
    \centering
    \includegraphics[width=\linewidth]{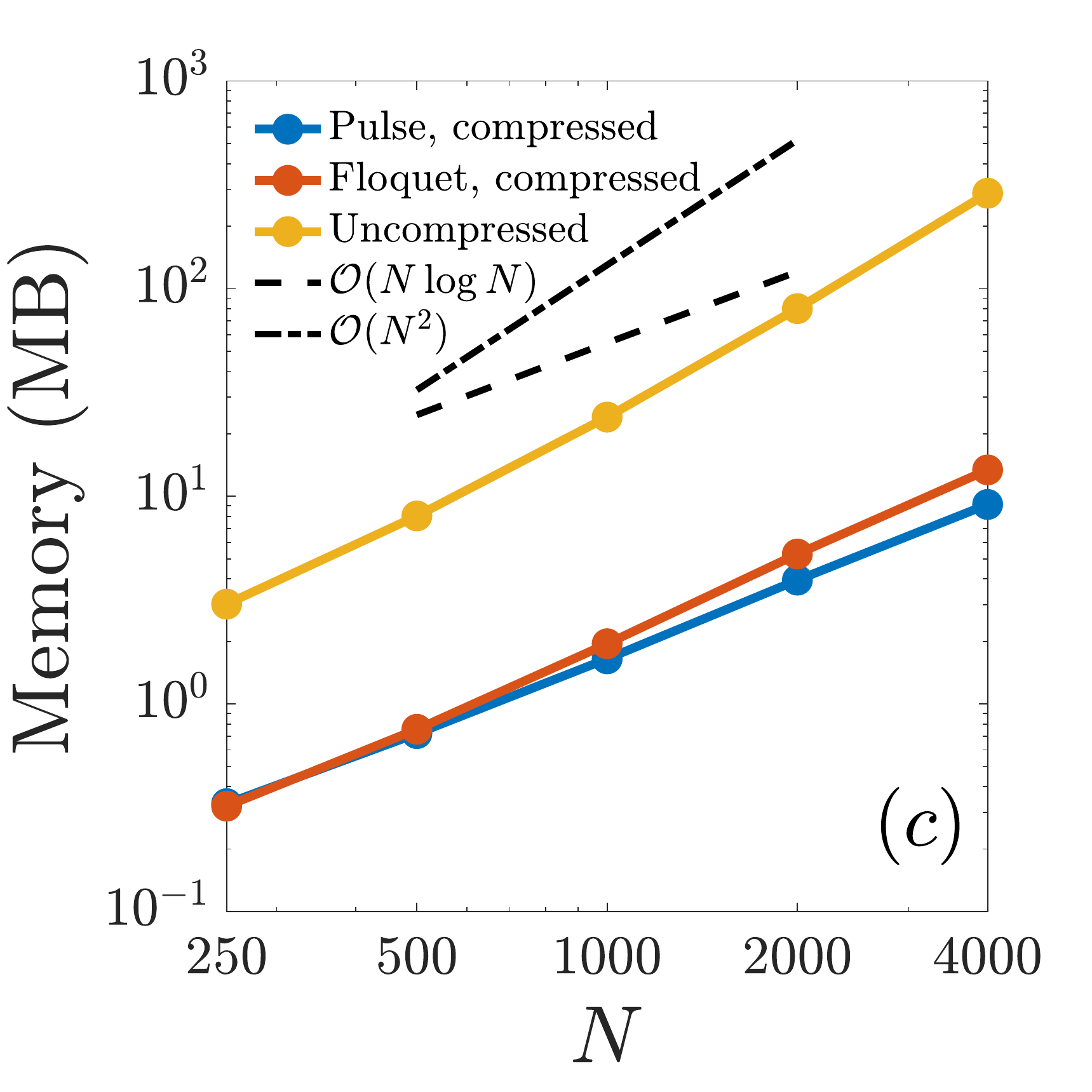}
    \captionlistentry{}
    \label{fig:nvmemoff_gw}
  \end{subfigure}
  \begin{subfigure}[t]{.23\textwidth}
    \centering
    \includegraphics[width=\linewidth]{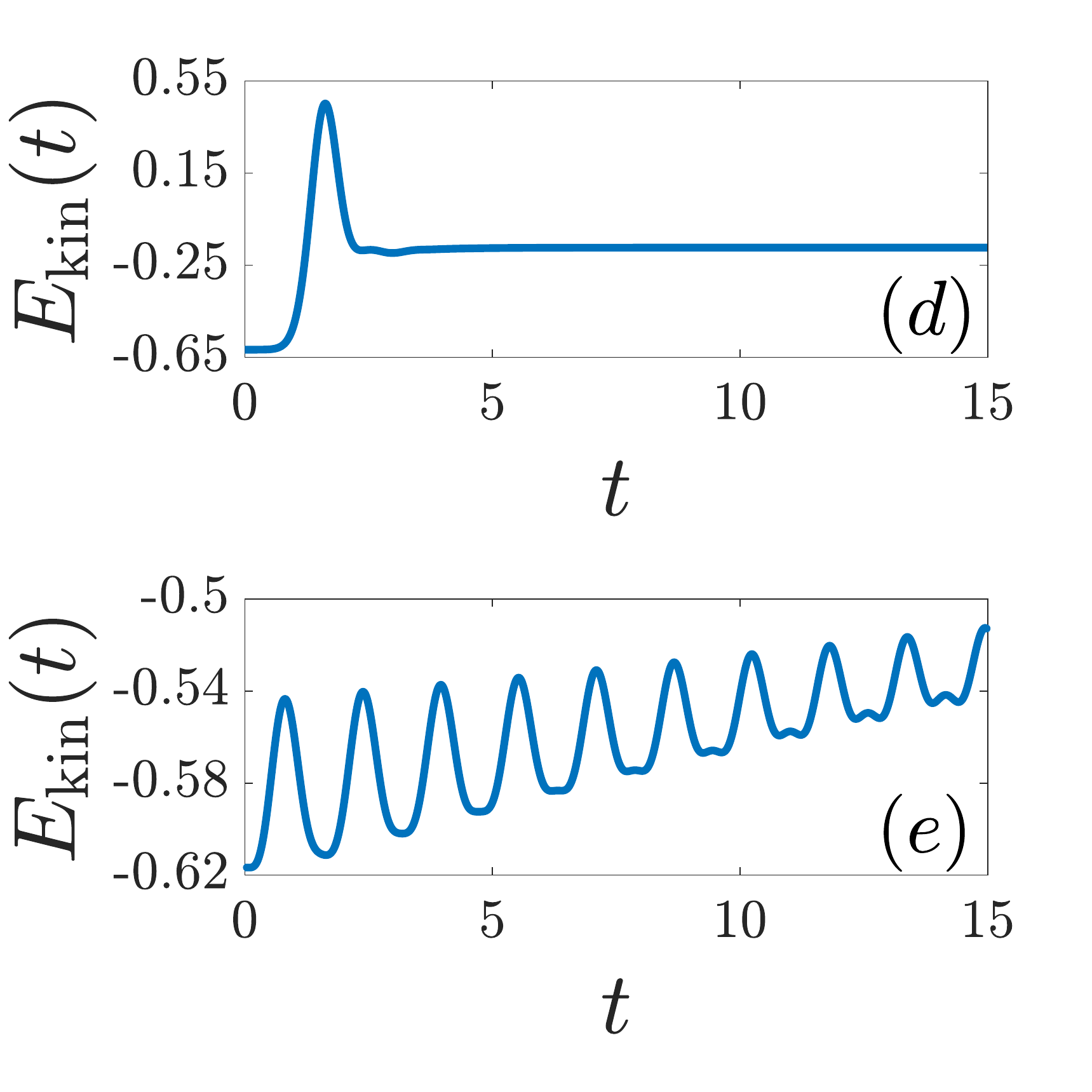}
    \captionlistentry{}
    \label{fig:gwke1}
    \captionlistentry{}
    \label{fig:gwke2}
  \end{subfigure}
  \vspace{-3ex}

  \caption{Ranks and memory usage for the GW examples with 
  increasing propagation times $T$ corresponding to $N$ time steps of
  fixed size $\Delta t$, computed offline from solutions obtained using
  the NESSi library.
  Panels (a) and (b) give $\varepsilon$-ranks for the pulse and ramp
  examples, respectively, with $\varepsilon = 10^{-4}$. Panel (c) gives
  the total memory usage in each case using compressed and direct storage.
  Panels (d) and (e) show the time evolution of the kinetic energy for the pump and
  Floquet examples, respectively.}

\label{fig:gwoff}
\end{figure}

The time evolution of the
kinetic energy $E_{\text{kin}}(t)=\frac{1}{N}\sum_k \epsilon(k) \langle
c_k^{\dagger} c_k \rangle(t)$ is shown for the pulse excitation and the
periodic driving in Figs. \ref{fig:gwke1} and \ref{fig:gwke2},
respectively. In the pulse excitation, the kinetic energy is
transiently enhanced during the pulse and then quickly approaches the
long-time limit which, as we are considering a closed system, is higher
than the equilibrium kinetic energy. In the periodically driven case,
the kinetic energy gradually grows toward the expected infinite
temperature state $E_{\text{kin}}=0$ as the system heats up. We note
that for readability, $E_{\text{kin}}(t)$ is plotted on a much shorter time interval than
that of our longer simulations.

We use the NESSi library~\cite{schuler2020} to solve the time-dependent
GW equations for these systems. 
We fix the time step $\Delta t = 0.01$ and the Matsubara time step $\Delta \tau = 0.04$, and compute
solutions $G_k(t,t')$ for $T
= 3.75,7.5,\ldots,60$, corresponding to $N = 250,500,\ldots,4000$, for both the pulse and Floquet examples. We then
measure the $\varepsilon$-ranks of all blocks in the compressed
representation for
$\varepsilon = 10^{-4}$,
and use these values, along with the number of directly stored matrix
entries, to compute the total memory required to store each Green's
function in
the compressed representation. The results are shown in Fig.
\ref{fig:gwoff}. The ranks grow slowly with $N$. Even for $N = 4000$, we observe a compression factor
of over $30$ for the pulse example and $20$ for the Floquet
example; this can be compared with a compression factor of approximately
$25$ for the Falicov-Kimball examples with $N=4096$. Of course, because
of the near linear scaling of the memory usage, the compression factors
will increase nearly linearly with $N$. As the two excitations represent very different physical
regimes, this experiment gives evidence of the broad applicability of
the HODLR compression technique.

\paragraph{Mott insulator within DMFT}
We next treat a strongly correlated Mott insulator. The description of
the Mott insulating phase of the Hubbard model requires a
nonperturbative approach, and we use the time-dependent DMFT
description~\cite{aoki2014}. For simplicity, we consider the Bethe
lattice self-consistency condition 
\beq{
  \Delta(t,t')=J(t) G_{\text{loc}}(t,t') J(t').
}
Here we have introduced the hybridization function $\Delta$ and the
local Green's function $G_{\text{loc}}=G_{ii}$. In the DMFT description,
the lattice problem is mapped to an effective impurity problem, and we
use the strong coupling expansion NCA as the impurity solver; see
Ref.~\onlinecite{eckstein2010} for details. To describe the electric
field on the Bethe lattice, we have followed the
prescription in
Refs.~\onlinecite{werner2017ultrafast,li2019long,werner2019entropy}.

As in the weak coupling case, the first excitation protocol is a
short electromagnetic pulse of the form \eqref{eq:pulse} with a single
cycle. We use $E_0 = 5$ and $\omega = 5$. The second is a periodic
driving of the form
\eqref{eq:floquet}, with $E_0^F = 1$ and $\omega_F = 5$. In both cases
we set $U = 6$ and $\beta = 20$. 

\begin{figure}[t]
  \centering
  \begin{subfigure}[t]{.23\textwidth}
    \centering
    \includegraphics[width=\linewidth]{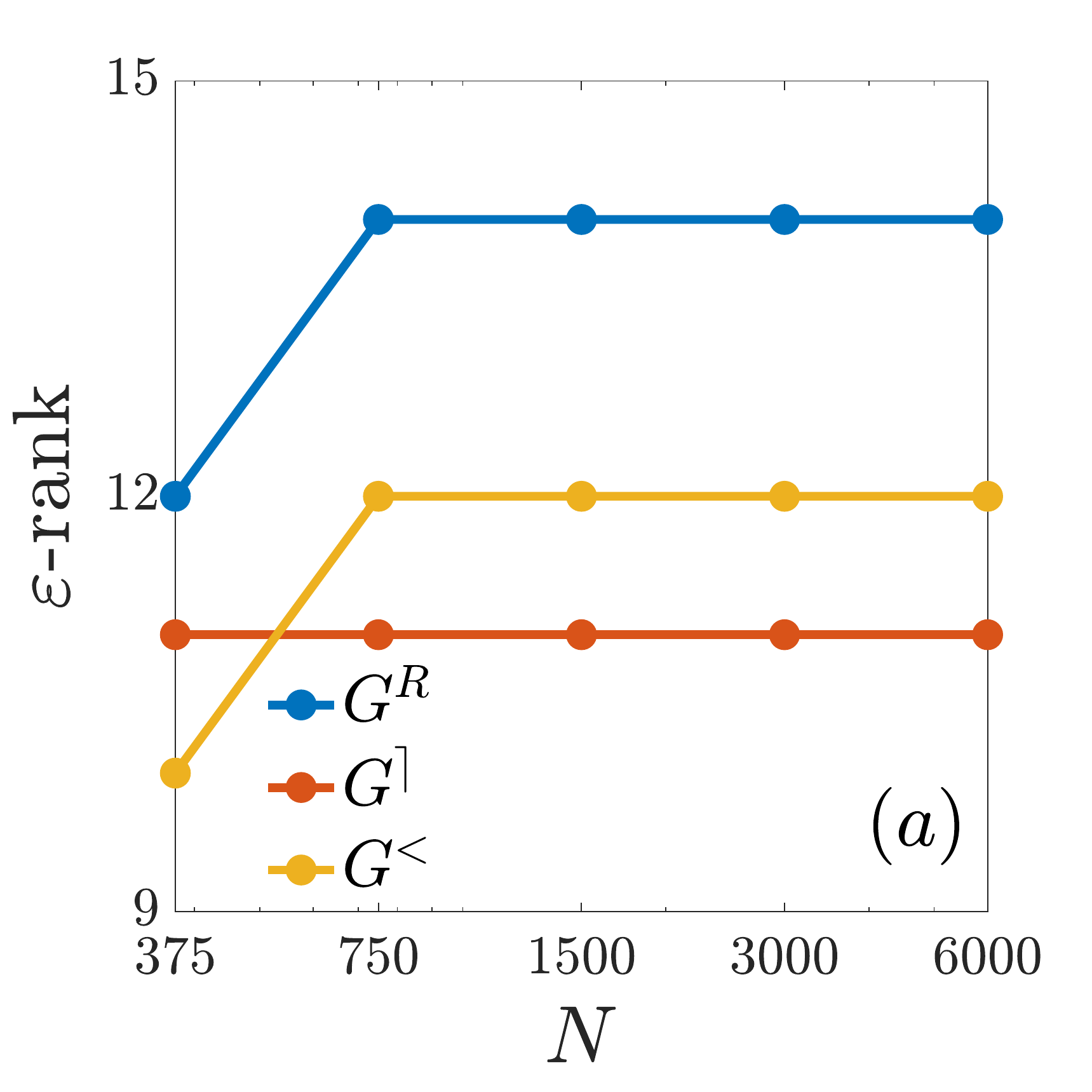}
    \captionlistentry{}
    \label{fig:nvrankoff_dmftpulse}
  \end{subfigure}
  \begin{subfigure}[t]{.23\textwidth}
    \centering
    \includegraphics[width=\linewidth]{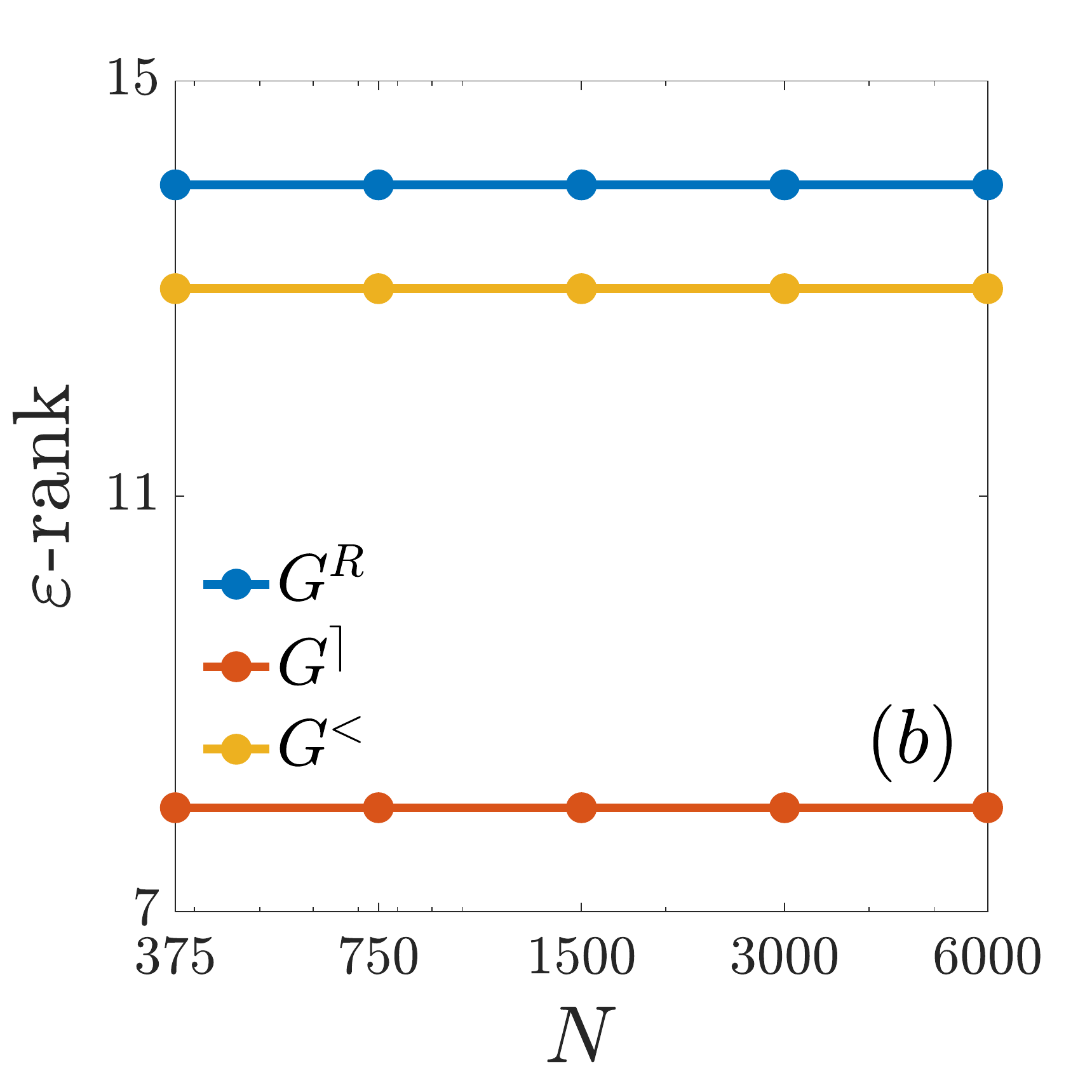}
    \captionlistentry{}
    \label{fig:nvrankoff_dmftfloq}
  \end{subfigure}
  \vspace{-3ex}

\begin{subfigure}[t]{.23\textwidth}
    \centering
    \includegraphics[width=\linewidth]{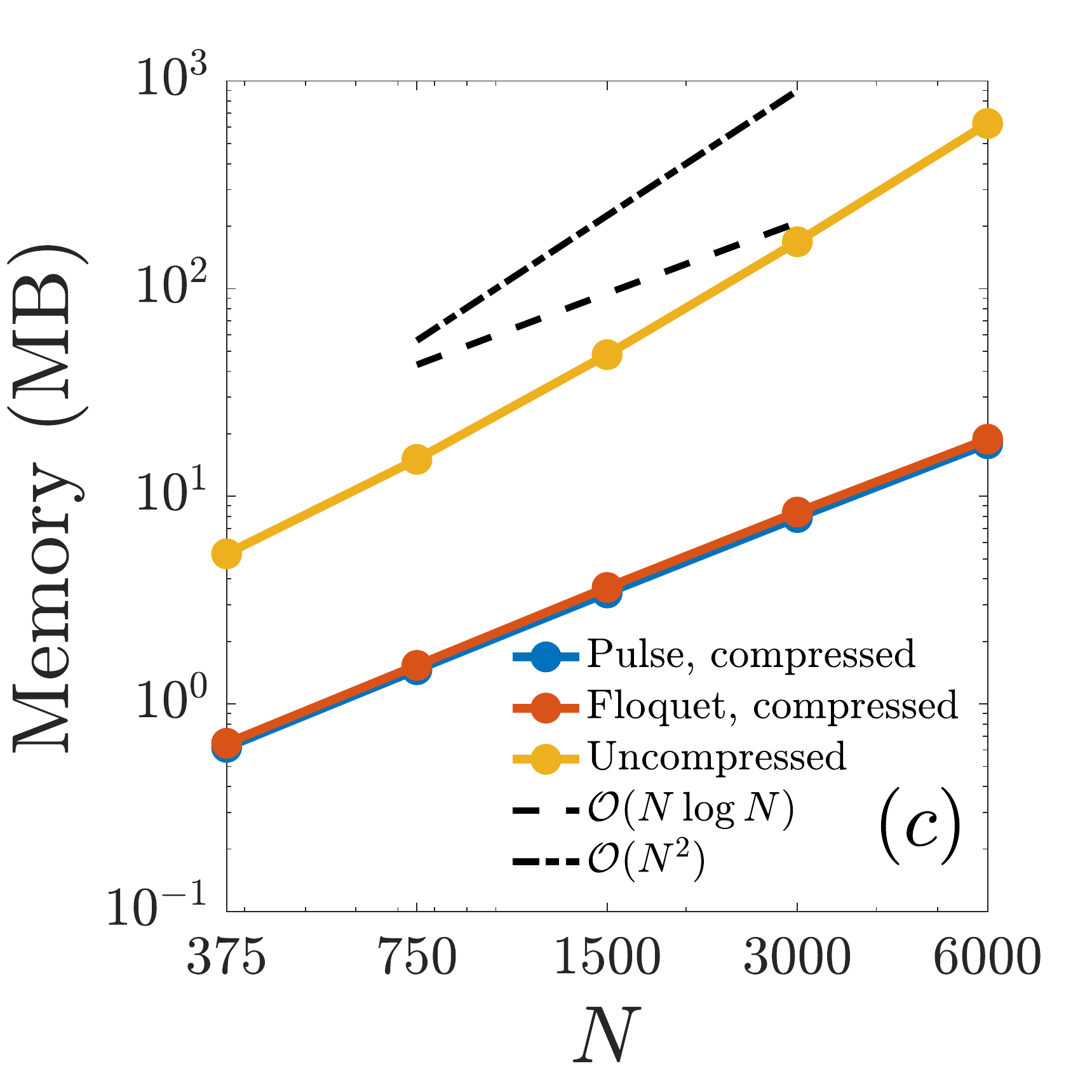}
    \captionlistentry{}
    \label{fig:nvmemoff_dmft}
  \end{subfigure}
  \begin{subfigure}[t]{.23\textwidth}
    \centering
    \includegraphics[width=\linewidth]{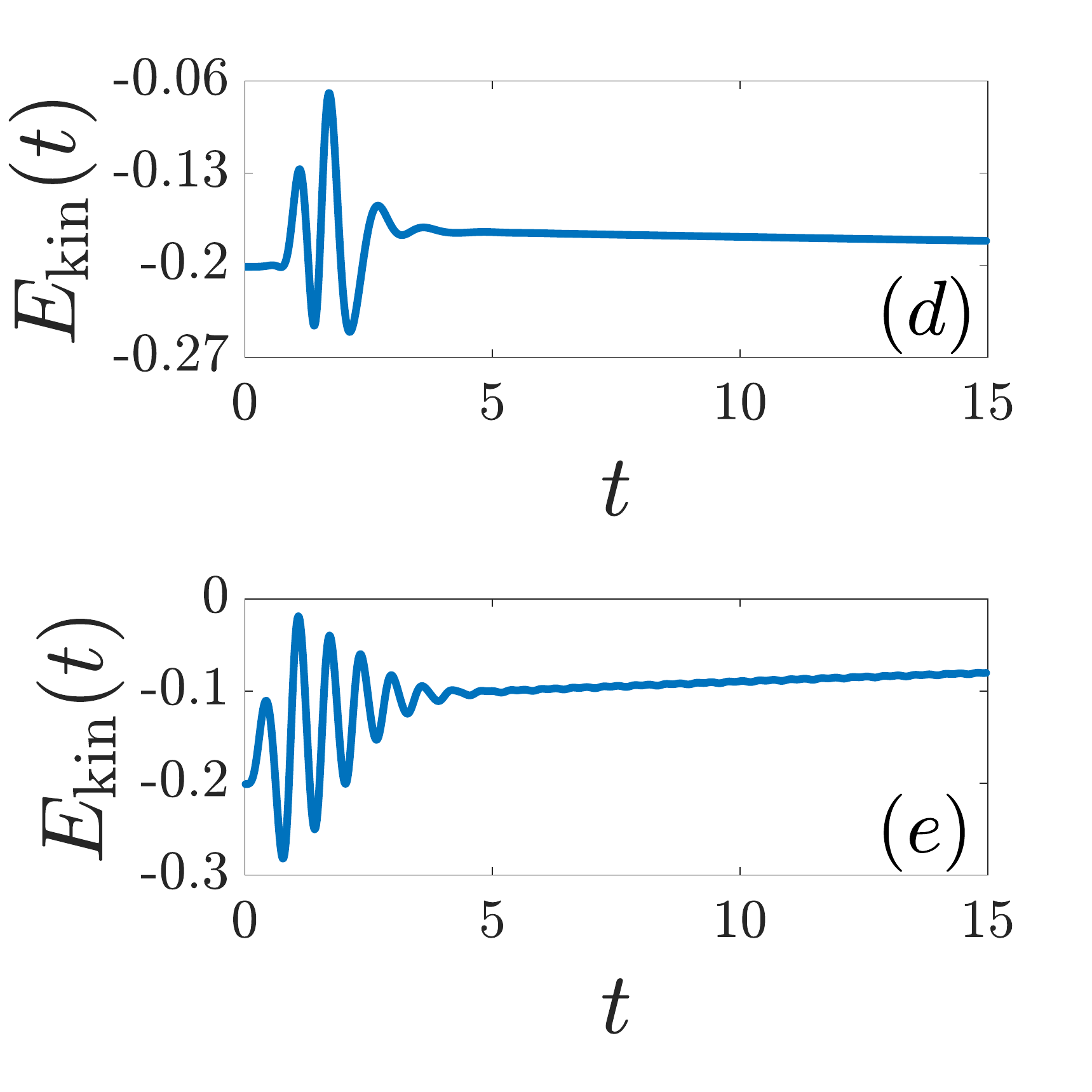}
    \captionlistentry{}
    \label{fig:dmftke1}
    \captionlistentry{}
    \label{fig:dmftke2}
  \end{subfigure}
  \vspace{-3ex}

  \caption{Ranks and memory usage for the DMFT examples with 
  increasing propagation times $T$ corresponding to $N$ time steps of
  fixed size $\Delta t$, computed offline from solutions obtained using
  the NESSi library.
  The plots are analogous to those shown for the GW examples in Fig.
  \ref{fig:gwoff}.}

\label{fig:dmftoff}
\end{figure}

The kinetic energy in the DMFT description is given by
$E_{\text{kin}}(t)=-2 \I \paren{\Delta*G_{\text{loc}}}^<(t,t)$. During the
pulse the kinetic energy, shown in Fig. \ref{fig:dmftke1}, increases and
then quickly relaxes to the long time limit. Despite the rather fast
relaxation to a nearly constant value of the kinetic energy,
computing this result is far from trivial as it requires integration over several
highly oscillatory auxiliary functions, often called pseudo-particle
propagators; see Ref.~\onlinecite{eckstein2010} for details. In the
Floquet example, the kinetic energy approaches the infinite
temperature state $E_{\text{kin}}=0$ in the long-time limit. While the initial dynamics show
strong oscillations, these are rapidly damped in the resonant regime;
see Refs.~\onlinecite{herrmann2018,peronaci2020,peronaci2018}. 

We fix the time step $\Delta t = 0.02$ and the Matsubara time step $\Delta \tau = 0.04$, and compute solutions
$G_k(t,t')$ for $T =
7.5,15,\ldots,120$, corresponding to $N = 375,750,\ldots,6000$, for both the pulse and Floquet examples.
We then measure $\varepsilon$-ranks for $\varepsilon = 10^{-4}$ and
memory usage as in the GW examples. The results are given in Fig.
\ref{fig:dmftoff}. The ranks remain constant as $N$
increases, and the memory scaling is consequently ideal. For $N = 6000$,
we obtain compression factors of over $30$ for both examples. Moreover,
we have confirmed that a similar degree of compression may be
obtained for the auxiliary pseudo-particle propagators. We note
that in these examples, the Green's functions and self energies exhibit rapid
decay in the off-diagonal direction, consistent with the
observed rank behavior, so both our method and methods based on
sparsity are applicable \cite{schuler2018,picano2020accelerated}. This demonstrates, in
addition, that matrices with rapid off-diagonal decay are in particular 
HODLR compressible.

\section{Conclusion}\label{Conclusion}

We have presented a numerical method to reduce the computational and
memory complexity of solving the
nonequilibrium Dyson equation by taking advantage of HODLR 
structures
of Green's functions and self energies observed in many physical systems. The method works
by building TSVD-based compressed representations of these functions on
the fly, and using them to reduce the cost of evaluating history
integrals. We have confirmed significant compressibility for various
models of interest, including instances of the Falicov-Kimball and
Hubbard models in different diagrammatic approximations. The accuracy of our method, compared with direct time
stepping methods, is controlled by the user, and
in particular does not involve any new modeling assumptions. Selection
of compression parameters is automatic, so our method may be used as a black box on
new systems with unknown structure.

This work suggests many important topics for future research, of which
we mention a few.

\begin{itemize}
    
\item Our method is compatible with more sophisticated discretization
  techniques, like high-order time stepping \cite{schuler2020}. While
    these do not by themselves change the computational and memory
    complexity of the solver, they yield a significant reduction in the
    constant associated with the scaling, and should be used in
    implementations. Also, in the equilibrium case, spectral methods
    and specialized basis representations have been 
    used to represent Green's functions with excellent efficiency, and their
    applicability in the nonequilibrium case has not yet been explored
    \cite{shinaoka2017,dong2020,li2020sparse,kananenka2016,otsuki2020sparse}.

\item A distinction must be made between automatic compression methods, like
the one we have described, and adaptive discretizations, which adjust
grids to increase resolution in certain regions of the solution.
Though significant technical challenges remain, combining compression and high-order discretizations with automatically
adaptive time stepping would enable the simulation of much more
sophisticated systems at longer propagation times, and we envision
such methods becoming the standard in the long term.

\item For systems amenable to HODLR compression, it remains to determine
whether this structure can be used to reduce other bottlenecks. In
particular, the evaluation of high-order self energy diagrams
involves a sequence of nested convolutions with potentially
structured operators.

\item The effectiveness of HODLR compression in solving the
  nonequilibrium Dyson equation is unsurprising, as various forms of
    hiearchical low rank compression are commonly used in scientific
    computing to compress integral operators with kernels that are
    smooth in the far field. However, since the equations are nonlinear,
    the degree of compressibility is difficult to analyze, and it
    remains to determine the limits of our approach. If HODLR
    compression is not applicable to some systems, it may still be
    possible to use similar ideas with other compression techniques from
    the numerical linear algebra and applied mathematics literature.
    Indeed, significant progress has been made over the last several
    decades on exploiting various types of data sparsity, especially in
    the context of partial differential equations and associated
    integral equations, and a significant opportunity remains to use
    these techniques in Green's function methods.

\end{itemize}

Lastly, there is an immediate opportunity to apply our method to
larger-scale physical simulations than were
previously possible. A full implementation in
the high-order time stepping code NESSI \cite{schuler2020} is forthcoming, and will be reported on at a later date.

\acknowledgements 
Simulations for the Hubbard model were carried out on the Rusty cluster
at the Flatiron Institute. For the solution of the strong coupling
Hubbard model, we have used a numerical library developed by Hugo U. R.
Strand and Martin Eckstein. We would like to acknowledge useful
discussions with Martin Eckstein, Yuta Murakami, Michael Sch\"uler, and Alex Barnett.  The Flatiron Institute is a division of the Simons Foundation.

\bibliography{kg_lrdyson}
\end{document}